\documentclass{JHEP3}

\usepackage{amsmath,amsfonts,amsthm,times}
\theoremstyle{plain}

\newtheorem{lem}{Lemma}

{\count255=\time\divide\count255 by 60
\xdef\hourmin{\number\count255}
\multiply\count255 by-60\advance\count255 by\time
\xdef\hourmin{\hourmin:\ifnum\count255<10 0\fi\the\count255}}
\def\ps@draft{\let\@mkboth\@gobbletwo
    \def\@oddhead{}
    \def\@oddfoot{\hbox to 7 cm{\tiny \versionno
       \hfil}\hskip -7cm\hfil\rm\thepage \hfil {\tiny\draftdate}}
    \def\@evenhead{}\let\@evenfoot\@oddfoot}
\def\draftdate{\number\month/\number\day/\number\year\ \ \ \hourmin }

\long\def\query#1{\hskip 0pt{\vadjust{\everypar={}\small\vtop to 0pt{\hbox{}%
     \vskip -13pt\rlap{\hbox to 49.0pc{\hfil{\vtop{\hsize=8pc\tolerance=6000%
     \hfuzz=.5pc\rightskip=0pt plus 3em\noindent#1}}}}\vss}}}}%
\newcommand\version[1] {\typeout{}\typeout{#1}\typeout{}
                   \vskip3mm \centerline{\fbox{{\tt DRAFT -- #1 -- }
                   {\small\draftdate}}} \vskip3mm}

\def\start{\documentclass[12pt]{article}
                   \usepackage{amssymb,amsfonts,amsmath,amscd,latexsym,epsf}   
\usepackage{showkeys}
                 \setlength{\textwidth}{17cm} \setlength{\textheight}{25.2cm}
 \hoffset -20mm \topmargin= -13mm \begin{document}\version\versionno
}

\def \ignore#1 { {} } 

\def \bs {\bigskip}

\def \Fig#1#2#3 {
\begin{figure}
\begin{center}
\scalebox{.6}{\includegraphics{#1.eps}}
\label{#1}
\end{center}
\caption{#3}
\end{figure}
}

\def \dis {\displaystyle}

\def \kz {{\rm KZ} }
\def \kzi {{\rm KZ}^{inv} }
\def \bpz {{\rm BPZ} }
\def \bpzi {{\rm BPZ}^{inv} }

\def \sp   { \ \ \ , \ \ \ }

\def \foint {\frac{1}{2i\pi}\oint }
\def \infint  {\int_{-\infty}^\infty }

\def \cst {{\rm constant}}

\def \half {\frac{1}{2}}
\def \halfpi {\frac{\pi}{2}}
\def \halfi {\frac{i}{2}}
\def \halfib {\frac{i}{2b^2}}
\def \halfb  {\frac{b}{2}}
\def \ip {i\frac{\pi}{2}}
\def \hib {\frac{1}{2b}}

\def \ap {\alpha'}
\def \ab {\bar{\alpha}}

\def \p {\partial}
\def \pp#1 {{\frac{\p}{\p #1}}}
\def \ppd#1 {{\frac{\p^2}{\p #1 ^2}} }

\def \om {\omega}
\def \g {\gamma}
\def \bg {\bar{\gamma}}
\def \vf {\varphi}
\def \s {\sigma}
\def \G {\Gamma}  
\def \e {\epsilon}
\def \a {\alpha}
\def \Ga {\Gamma_b}
\def \up {\Upsilon_b}
\def \t {\theta}

\def \rar {\rightarrow}
\def \lrar {\leftrightarrow}

\def \j  {-\half+iP}

\def \bz {\bar{z}}
\def \bw {\bar{w}}
\def \bx {\bar{x}}
\def \bp {\bar{\partial}}
\def \bq {\bar{q}}
\def \bL {\bar{L}}
\def \bJ {\bar{J}}

\def \Tr {{\rm Tr }\ }
\def \id {{\rm id } }

\def \Z {\mathbb{Z}}
\def \N {\mathbb{N}}
\def \R {\mathbb{R}}
\def \C {\mathbb{C}}
\def \S {\mathbb{S}}

\def \la {\left\langle}
\def \ra {\right\rangle}
\def \lla {\langle\langle}
\def \rra {\rangle\rangle}

\def \tq {\tilde{q}}
\def \tth {\tilde{\theta}}
\def \ttau {\tilde{\tau}}

\def \uq  {U_q(sl_2)}
\def \asl {\widehat{sl_2}}

\def \vt {\vartheta_1}

\def \Lrange {{\frac{Q}{2}+i\R_+}}
\def \Hrange {{-\half +i\R_+}}
\def \Hfull  {{-\half +i\R}}
\def \SLC    {SL(2,\C)}
\def \SLR    {SL(2,\R)}
\def \SLU    {SL(2,\R)/U(1)}
\def \H      {H_3^+}
\def \tR     {\tilde{R}}
\def \FL#1#2#3#4#5#6{
F^L_{#5\, #6}\left[\begin{array}{cc} 
#3 & #2 \\ 
#4 & #1
\end{array}\right]}
\def \FM#1#2#3#4#5#6{
F_{#5\, #6}\left[\begin{array}{cc} 
#3 & #2 \\ 
#4 & #1
\end{array}\right]}
\def \FH#1#2#3#4#5#6{
F^{H_3}_{#5\, #6}\left[\begin{array}{cc} 
#3 & #2 \\ 
#4 & #1
\end{array}\right]}
\def \Fh#1#2#3#4#5#6{
F^{h}_{#5\, #6}\left[\begin{array}{cc} 
#3 & #2 \\ 
#4 & #1
\end{array}\right]}
\def \GL#1#2#3#4#5#6{
G^{L}_{#5\, #6}\left[\begin{array}{cc} 
#3 & #2 \\ 
#4 & #1
\end{array}\right]}
\def \CG#1#2#3#4#5#6{ \left(\begin{array}{ccc} #1 & #2 & #3 \\ #4 & #5
    & #6 \end{array} \right) }
\def \vir { \ , \ }
\def \F32#1#2#3#4#5#6{{} _3F_2\left(\left.\begin{array}{c}#1 \vir #2 \vir
    #3 \\ #4\vir #5
    \end{array} \right| #6 \right) }

\def\vect#1#2{ \left( \begin{array}{c} #1 \\ #2  \end{array} \right)}

\def \sgn{ {\rm sgn} }
\def \exch{1\lrar 3,2\lrar 4,\sigma \lrar \tau}

\def \bea {\begin{eqnarray}}
\def \eea {\end{eqnarray}}
\def \bee {\begin{eqnarray*}}
\def \eee {\end{eqnarray*}}
\def \nn  {\nonumber}
\def \sst {\scriptscriptstyle}

\newcommand {\citer}[2]{\begin{flushright}{\small{[\ #1,\ \tcm{\it
          #2}\ ]}}
\end{flushright}}

\newcommand {\slidetitle}[1]{ \large \begin{center} { #1} \\
  \end{center} \normalsize \vspace{0.8cm}  }

\newcommand {\tcb}[1]{#1}
\newcommand {\tcr}[1]{#1}
\newcommand {\tcg}[1]{#1}
\newcommand {\tcm}[1]{#1}
\newcommand {\tcc}[1]{#1}
\newcommand {\tcy}[1]{#1}
\newcommand {\tcnoir}[1]{#1}

\def \FF { {} _2{\cal F}_1 } 
\def \zbz { {\footnotesize \begin{array}{c} z_1\dots z_n \vspace{-1.3mm}\\
\bz_1\dots \bz_n
\end{array}} } 
\def \yby { {\footnotesize \begin{array}{c} y_1\dots y_{n-2} \vspace{-1.3mm}\\
 \bar{y}_1\dots
\bar{y}_{n-2} \end{array}} }
\def \mubmu { {\footnotesize \begin{array}{c} \mu_1\dots \mu_{n} \vspace{-1.3mm}\\
 \bar{\mu}_1\dots
\bar{\mu}_{n} \end{array}} }
\def \mubmuf { {\footnotesize \begin{array}{c} \mu_1\dots \mu_{4} \vspace{-1.3mm}\\
 \bar{\mu}_1\dots
\bar{\mu}_{4} \end{array}} }

\def \rml {L}
\def \rmh {H}

\title{$H_3^+$-WZNW correlators from Liouville theory }

\author{ Sylvain Ribault\,$^1$\, and \,Joerg Teschner\,$^2$
\vspace{5mm}
\\
\!\!\!$^1${\it  King's College London
 \\  Department of Mathematics
\\ Strand, London WC2R 2LS, United Kingdom }
\\ {\tt ribault@mth.kcl.ac.uk }
\vspace{3mm}
\\
\!\!\!$^2${\it Institut f\"ur theoretische Physik
\\ Freie Universit\"at Berlin
\\ Arnimallee 14, 14195 Berlin, Germany }
\\ {\tt teschner@physik.fu-berlin.de }
}

\abstract{
We prove that arbitrary correlation functions of the $\H$-WZNW model on a
sphere have a simple expression in terms of Liouville theory
correlation functions. This is based on the correspondence between the
KZ and BPZ equations,
and on relations between the structure constants of Liouville theory
and the $\H$-WZNW model. In the critical level limit, these results imply a
direct link between eigenvectors of the Gaudin Hamiltonians and the
problem of uniformization of Riemann surfaces. We also 
present an expression for correlation functions of the 
$SL(2)/U(1)$ gauged WZNW model 
in terms of correlation functions in Liouville theory.
}

\preprint{ 
\hepth{0502048}\\ KCL-MTH-05-01 \\
 }

\begin{document}

\section{Introduction}

Liouville theory and the $H_3^+$-WZNW model 
(henceforth mostly abbreviated as $H_3^+$ model)
are two conformal field theories which have played an important 
r\^ole in many recent studies of noncritical string theories, 
two-dimensional quantum gravity and the AdS/CFT correspondence in the
case of the three-dimensional Anti-de Sitter space. These two models
are examples of
conformal field theories with continuous sets of primary fields
which are not obtained from free field theories in any 
simple way.
 One may view Liouville theory and the $H_3^+$ model
as ``noncompact'' counterparts of minimal models and the 
$SU(2)$ Wess--Zumino--Novikov--Witten model respectively. These and other observations
form the basis for our expectation that 
Liouville theory and the $H_3^+$ model could play
a similar r\^ole in the development of non rational conformal 
field theories as the minimal and WZNW models have played 
in making rational conformal field theories a powerful 
theoretical tool with various applications. It is also
worth noting that a certain number of conformal field theories
with high string-theoretical relevance such as the 
$SL(2)/U(1)$ Euclidean black hole model and 
the N=2 Liouville theory can be obtained from the 
(supersymmetrized) $H_3^+$ model by means of 
simple coset constructions.

The search for relations between Liouville theory and the 
Minkowskian counterpart of the $H_3^+$ model, the 
$SL(2,{\mathbb R})$ WZNW model, has a long tradition, stimulated by
\cite{Pol,KPZ}. Originally it was hoped that the 
$SL(2,{\mathbb R})$ symmetry of the $SL(2,{\mathbb R})$ WZNW model 
would make it easier to solve this model first, from which 
quantum Liouville theory would be obtained by means of a quantum 
version of Hamiltonian reduction \cite{FWBFR}.

Our point of view will be quite the opposite one. At the moment,
it is Liouville theory which is by far the best understood
example of an interacting conformal field theory with
a continuous set of primary fields, see \cite{TR,TT,Na} and 
references therein. The $H_3^+$  model
is only reasonably well understood on punctured Riemann surfaces
of genus zero \cite{Ga,TeH1,TeH2,TeH3}. 
Not as well understood is the $H_3^+$ model
on Riemann surfaces with boundaries. This is probably due to the
fact that in contrast to the case of Liouville theory we do
not understand the chiral bootstrap of the $H_3^+$ model 
properly yet\footnote{It may well be that a generalization of the
formalism usually called chiral bootstrap is needed in this case since
the representations which generate the spectrum of the $H_3^+$ model 
do not exhibit chiral factorization in the usual sense.}.
Even less understood is the  $SL(2,{\mathbb R})$ WZNW model. Despite some
important progress \cite{MO}, we do not control the definition
and the analytic properties of the correlation functions with 
more than three field insertions yet.

Our main motivation for seeking relations between Liouville theory on the
one hand, and the $H_3^+$ model on the other 
hand is therefore the hope that we can use knowledge from 
Liouville theory to improve upon our understanding of the 
$H_3^+$ and $SL(2,{\mathbb R})$ models. In the present paper
we shall present an explicit formula relating arbitrary
correlation functions of Liouville theory and the $H_3^+$ model
on punctured Riemann spheres. The simplicity of this relation
makes us view the $H_3^+$ model and its cosets like
the $SL(2)/U(1)$ gauged WZNW model as ordinary Liouville theory in disguise.
We plan to discuss
similar relations between the conformal blocks of these theories in subsequent
publications, and we hope that this will also allow us to make 
progress on the boundary problem.
We also believe that similar techniques 
can be used to construct the correlation 
functions of the $SL(2,{\mathbb R})$ model.

Another application that we have in mind is the investigation of 
the critical level limits of the $H_3^+$ and $SL(2)/U(1)$ models.
It was conjectured by V. Fateev and the brothers Zamolodchikov that 
the $SL(2)/U(1)$ gauged WZNW model is dual to sine-Liouville theory. The 
supersymmetric counterpart of this duality can be seen as mirror 
symmetry \cite{HK}. These dualities should manifest 
themselves most clearly in the critical level
limit, where the 
sine-Liouville and the N=2 Liouville actions 
would provide weakly coupled descriptions of the 
respective models. It is possible to 
construct correlation functions in all these models from the 
$H_3^+$ model \cite{Ga,GK}. Given these relations it
is natural to investigate if the critical level
limit may also be seen as a dual weak coupling limit in the 
case of the $H_3^+$ model. Results about the 
critical level limit of the $H_3^+$ model could then help us 
to deepen our understanding of the above-mentioned 
non-trivial dualities. Some steps in this direction
will be taken in the present paper by relating the critical 
level limit of the $H_3^+$ model to the semiclassical
limit of ordinary bosonic Liouville theory.

\paragraph{Plan of the paper. } 

We start by collecting those results about 
Liouville theory and the $\H$ model which will be relevant for 
us. The main novelty in this section will be to introduce a new 
basis for the space of primary fields 
in the $\H$ model which will simplify the analysis  considerably.

We then state our main result, equation (\ref{main}). This is the
expression of arbitrary $n$-point correlators of the $\H$ model on the
sphere in terms of Liouville correlators. We prove it by induction on
$n$. An important ingredient is a correspondence between the
Knizhnik--Zamolodchikov and Belavin--Polyakov--Zamolodchikov equations
in a version due to Stoyanovsky, which we review. We also derive
useful relations between the three-point structure constants of the
models. Next we compare our main result with the original version of
the KZ-BPZ correspondence due to Fateev and Zamolodchikov\footnote{
Other works on relations between the KZ equations and the
null vector decoupling equations include \cite{CF,FGPP,PRY}. However,
in these cases only the admissible representations of the
$sl(2)$ current algebra were considered. The connection between
these results and the
correspondence discovered by Stoyanovsky is not obvious to us.
The latter correspondence is related to and motivated by
earlier results on relations between the Gaudin model,
its solutions via the Bethe ansatz and the KZ equations,
see \cite{B,BF,FFR,Fr}.}. We also
indicate how correlators of the $\SLU$ coset model are related to
Liouville correlators.

The last section is devoted to the study of the critical level limit
$k\downarrow 2$. In this limit, the KZ equations reduce to the
eigenvector equations for the Gaudin Hamiltonians. The corresponding limit
in Liouville theory is known to be related to the uniformization of Riemann
surfaces with conical singularities. We use known results for
this problem in order to build particular eigenvectors of the Gaudin
Hamiltonians. The corresponding eigenvalues are given by the so-called
accessory parameters of the classical Liouville solution.

\section{Review of Liouville theory and the $\H$ model }

\subsection{Liouville theory }

Liouville theory is defined classically by the action
\begin{equation}\label{L-act}
S^L\,=\,\frac{1}{\pi}
\int d^2w\,\big(|\partial_w\phi|^2+\mu_L e^{2b\phi}\big)\,.
\end{equation}
Liouville theory is a conformal field theory \cite{TR,TT} whose
left- and right-moving Virasoro algebras are respectively generated by the modes of
the stress-energy tensors $T(w)=-(\p \phi)^2+Q\p^2 \phi$ and
$\bar{T}(\bw)=-(\bar{\p} \phi)^2 +Q \bar{\p}^2 \phi$.
The central charge $c$ and the background charge $Q$ are
expressed in terms of the parameter $b$ via \footnote{The value of $Q$
  follows from the requirement that the Liouville interaction term
  $\mu_L e^{2b\phi}$ be exactly marginal.}
\begin{equation}
c=1+6Q^2,\qquad Q\equiv b+b^{-1}\,.
\end{equation}
The primary fields of Liouville theory are denoted
as $V_{\a}(z)$ and reduce to the exponential fields $e^{2\a\phi}$ in
the classical limit $b\rightarrow 0$.
The parameter $\a$ may take arbitrary 
complex values, but the subset $\a \in 
\frac{Q}{2}+i\R$ is distinguished
by the fact that the action of $V_{\a}(z)$ on the
vacuum state $|0\rangle$ creates
delta-function normalizable states. 
The primary field $V_{\a}(z)$ has conformal
weight 
\begin{equation}
\Delta_\a=\a (Q-\a)\,.
\end{equation}

\subsubsection{Correlation functions}

The theory is fully characterized by 
the $n$-point correlators on a sphere :
\begin{equation}\label{L-npt}
\Omega^L_n(z_n,\dots,z_1|\a_n,\dots,\a_1)\,
=\,\big\langle\, V_{\a_n}(z_n)\cdots V_{\a_1}(z_1)\,\big
\rangle .
\end{equation}
Power series representations for $\Omega^L_n$ may be obtained 
by
using the operator product expansion
\begin{equation}\label{L-OPE}
\begin{aligned}
V_{\a_2}(z_2)& V_{\a_1}(z_1)\,=\, \frac{1}{2\pi}\int_{\frac{Q}{2}+i\R_+} d\a_3\;
\big\langle\, V_{\a_3}(\infty) V_{\a_2}(z_2-z_1)V_{\a_1}(0)\,\big
\rangle\\
&\hspace{5cm}\times\big(V_{Q-\a_3}(z_1)
+({\rm descendants})\big)\,,
\end{aligned}
\end{equation}
where  
$V_{\a_3}(\infty)$ is defined so that correlation functions involving
it are finite, 
\begin{equation}\label{inftynot}
V_{\a_3}(\infty)\,=\,\lim_{z\rar \infty} |z|^{4\Delta_\a} \langle 0| V_\a
(z) \,.
\end{equation}
The descendant contributions in (\ref{L-OPE}) are fully determined 
by conformal covariance of the OPE, however these contributions are
not known explicitly. In principle, the construction of general
$n$-point correlation functions therefore boils down to the 
construction of the 2- and 3-point functions, which are 
of the form 
\begin{align}
\langle\,V_{\a_2}(z_2) V_{\a_1}(z_1)\,\rangle\,=\,&2\pi(
\delta(Q-\a_2-\a_1)+R^{ L}(\a_1)\delta(\a_2-\a_1)) 
\nn \\ & \times {|z_2-z_1|^{-4\Delta_{\a_1}}}\,,  \label{L-2pt}\\
\langle\,V_{\a_3}(z_3) V_{\a_2}(z_2) V_{\a_1}(z_1)\,\rangle\,=\,& 
|z_3-z_2|^{-2\Delta_{32}^1}|z_3-z_1|^{-2\Delta_{31}^2}
|z_3-z_2|^{-2\Delta_{21}^3}\nn \\ &\times 
C^{ L}(\a_3,\a_2,\a_1), \label{L-3pt}
\end{align}
where $\Delta_{rs}^t\equiv\Delta_{\a_r}+\Delta_{\a_s}-\Delta_{\a_t}$.
The following expressions for the structure functions $C^{L}$
and $R^L(\a)$ 
were proposed in \cite{DO,ZZ}:
\begin{align}\label{L-3c}
C^{L}(\a_3,\a_2,\a_1) =&\;  
\frac{\left[\pi\mu_{L}\gamma(b^2)b^{2-2b^2}
\right]^{b^{-1}(Q-\a_1-\a_2-\a_3)} }{
\Upsilon_b(\alpha_1+\alpha_2+\alpha_3-Q)
} \\
& \;\;
\times\frac{\Upsilon_b'(0)\Upsilon_b(2\alpha_1)\Upsilon_b(2\alpha_2)
\Upsilon_b(2\alpha_3)}{\Upsilon_b(\alpha_1+\alpha_2-\alpha_3)
\Upsilon_b(\alpha_2+\alpha_3-\alpha_1)
\Upsilon_b(\alpha_3+\alpha_1-\alpha_2)}\nn \\
R^{\rml}(\a)= (\pi \mu_{\rml} 
& \gamma(b^2)  )^{\frac{Q-2\a}{b}}
\frac{\Gamma(1+b(2\a-Q))}{\Gamma(1-b(2\a-Q))}
\frac{\Gamma(1+b^{-1}(2\a-Q))}{\Gamma(1-b^{-1}(2\a-Q))}
\label{L-2c}
\end{align}
The quantity $R^L(\a)$ is the reflection coefficient of Liouville
theory. It appears in the reflection relation,
\bea
\label{refliou}
V_\a=R(\a)V_{Q-\a}.
\eea
The existence of such a relation explains why the physical spectrum is
generated by the states with $\a\in \frac{Q}{2}+i\R_+$ instead of
$\a\in \frac{Q}{2}+i\R$, a fact we have already used in writing the
OPE (\ref{L-OPE}).

Definition and relevant properties of the special
function $\Upsilon_b(x)$ are recalled in Appendix \ref{apup}.
Consistency of the conformal field theory characterized by
equations of (\ref{L-OPE})-(\ref{L-2c}) was
proven in \cite{TT}. There
is ample evidence \cite{ZZ,TR,Th} that this conformal field theory
is a quantization of the classical theory described by the
action (\ref{L-act}) with $V_{\a}(z)$ being the quantized
counterpart of $e^{2\a\phi(z)}$.

\subsubsection{Degenerate fields and BPZ equations}

The Liouville correlators $\Omega_n^L$
(\ref{L-npt}) are a priori defined only for physical values of the
momenta, $\a_i \in \frac{Q}{2} + i\R$. However, they
possess a meromorphic continuation to arbitrary complex momenta
$ \a_i \in\C $ \cite{TR}. In particular, we may specialize to correlators
involving fields with momentum $\a=-\hib$ which 
correspond to the degenerate Verma module for the Virasoro algebra
with Kac labels $(1,2)$, 
\begin{equation}\label{L-nmpt}
\Omega^L_{n|m}\,\equiv\,
\big\langle\, V_{\a_n}(z_n)\cdots V_{\a_1}(z_1)\,
V_{-\frac{1}{2b}}(y_m)
\cdots
V_{-\frac{1}{2b}}(y_1)\,
\big\rangle\,.
\end{equation}
Each degenerate field $V_{-\hib}(y_r)$ gives rise to a
Belavin--Polyakov--Zamolodchikov (BPZ) differential equation for the
correlator $\Omega^L_{n|m}$,
\begin{equation}\label{BPZeqns}
{\rm D}^{\rm\sst BPZ}_r\,\Omega^L_{n|m}\,=\,0\,,
\end{equation}
where the differential operator ${\rm D}^{\rm\sst BPZ}_r$ is defined 
as
\begin{equation}
\begin{aligned}
{\rm D}^{\rm\sst BPZ}_r\,=\,b^2\ppd{y_r} +
\sum_{s\neq r}\biggl(\frac{1}{y_r-y_{s}} & \pp{y_{s}}
+\frac{\Delta_{-\frac{1}{2b}}}{(y_r-y_{s})^2} \biggr)\\ &
+\sum_s\left(
\frac{1}{y_r-z_s} \pp{z_s} +\frac{\Delta_{\a_s}}{(y_r-z_s)^2}\right)\,.
\end{aligned}
\end{equation}
The differential equations (\ref{BPZeqns}) express the decoupling
of the singular vector in the Verma module 
with highest weight $\Delta_{-1/2b}$.
The equations
(\ref{BPZeqns}) imply in particular that the 
OPE $V_{-\frac{1}{2b}}(y)V_{\a}(z)$ takes the simple form
\begin{equation}\label{degOPE}
V_{-\frac{1}{2b}}(y)V_{\a}(z)\,=\, \sum_{\eta=\pm}\,
|y-z|^{\eta(2\a-Q)+Q}C^L_{\eta}(\a)
V_{\a-\frac{\eta}{2b}}(z)+({\rm descendants}) \,,
\end{equation}
where the OPE coefficients are
\begin{equation}\begin{aligned}
C^L_-(\a)\,=\,& R^L(\a)R^L(Q-\a-\hib)=(\pi
\mu_L\gamma(b^2))^\frac{1}{b^2} b^{-4}
\frac{\gamma(2b^{-1}\a-1-b^{-2})}{\gamma(2b^{-1}\a)}
\,,\\
C^L_+(\a)\,=\,& 1\,.
\end{aligned}
\end{equation}

\subsection{The $\H$ model } 

The classical $\H$ model is defined by the following 
action:
\begin{equation}
S^H\,=\,k \int d^2 z\; 
\left(\p \phi \bp \phi +e^{2\phi} \p \g \bp \bg \right)\,.
\label{H-act}
\end{equation}
The quantum $\H$-model \cite{TeH1}-\cite{TeH3} 
is a conformal field theory that has the current
algebra 
$(\widehat{\mathfrak s\mathfrak l}_2)_k
\times(\widehat{\mathfrak s\mathfrak l}_2)_k$
as a symmetry. The corresponding currents will be denoted
by $J^a(z)$ and $\bar{J}^a(\bz)$ respectively.

The primary fields of the $\H$-model are usually parametrized
as $\Phi^j(x|z)$, $z\in\C$, $x\in\C$, and are the quantized
counterparts of the following functions on $\H$ :
\begin{equation}
\Phi^{j}_{\rm\sst cl}(x|z)\,\equiv\,
\frac{2j+1}{\pi}\big(|\gamma-x|^2e^{\phi}+e^{-\phi}\big)^{2j}\,.
\end{equation}
The conformal weight of the primary field $\Phi^j(x|z)$ is 
\bea
\Delta_j=-b^2j(j+1)=-\frac{j(j+1)}{k-2} \,,
\eea
where we have introduced the notation $b^2=1/(k-2)$. This notation
anticipates the relation with the $b$ parameter of Liouville theory.

Our review of the $\H$ model and our analysis of its correspondence
with Liouville theory will be greatly simplified by the introduction
of a new set of primary fields $\Phi^j(\mu|z)$, which are defined from
$\Phi^j(x|z)$ by a change of basis within each representation of spin $j$,
\bea
\Phi^j(\mu|z)=\frac{1}{\pi}
|\mu|^{2j+2} \int_\C d^2x\ e^{\mu x-\bar{\mu}\bar{x}} \Phi^j(x|z) \,.
\label{phimu}
\eea
The reader may find the usual $x$-basis quantities, and
the derivation of our new $\mu$-basis quantities, in Appendix
\ref{H-xvar}. 

The fields $\Phi^j(\mu|z)$ are characterized by 
the following operator product expansions with the currents:
\begin{equation}
J^a(w)\Phi^j(\mu|z)\,\sim\,\frac{1}{w-z}{\rm D}^a \Phi^j(\mu|z),
\quad
\bar{J}^a(\bw)\Phi^j(x|z)\,\sim\,\frac{1}{\bw-\bz}
\bar{\rm D}^a\Phi^j(x|z)\,.
\end{equation}
The differential operators ${\rm D}^a_j$
are defined as
\begin{equation}
{\rm D}^-=\mu ,\quad
{\rm D}^0=-\mu \p_\mu ,\quad
{\rm D}^+=\mu \p^2_\mu-\frac{j(j+1)}{\mu},
\end{equation}
and the quadratic Casimir of the corresponding representation is 
\bea
{\rm D}^a{\rm D}^a= ({\rm D}^0)^2-\half({\rm D}^+{\rm D}^-+{\rm
  D}^-{\rm D}^+)=j(j+1)\,. 
\label{casimir}
\eea
The
operators $\bar{\rm D}^a_j$ are the complex conjugates of 
${\rm D}^a_j$. 

We may consider arbitrary 
complex values of $j$, but the values  
$j\in -\frac{1}{2}+i\R$ are distinguished
by the fact that the action of $\Phi^j(\mu|z)$ on the
vacuum $|0\rangle$ creates
delta-function normalizable states, which generate the physical spectrum of the
$\H$ model. These values of $j$ are those which appear in the
decomposition of the classical limit $L^2(\H)$ of the spectrum. This
classical limit is defined by $k\rar \infty$. 

\subsubsection{Correlation functions}

The  $\H$-model is fully characterized by the 
$n$-point correlation functions on a sphere :
\begin{equation}
\Omega^{\rmh\sst}_n(\mu_n,\dots,\mu_1|z_n,\dots,z_1|j_n,\dots,j_1)\,
=\,\big\langle\, \Phi^{j_n}(\mu_n|z_n) \cdots
\Phi^{j_1}(\mu_1|z_1) \,\big\rangle\,.
\end{equation}
These correlation functions may be constructed by using the
operator product expansion
\begin{equation}\label{H-OPE}
\begin{aligned}
\Phi^{j_2}(\mu_2|z_2) & 
\Phi^{j_1}(\mu_1|z_1)\,=\,\\ 
& =\,\int_{-\half+i\R_+}dj_3\int \frac{d^2 \mu_3}{|\mu_3|^2} \;
\big\langle\, \Phi^{j_3}(\mu_3|\infty) \Phi^{j_2}(\mu_2|z_2-z_1)
\Phi^{j_1}(\mu_1|0) \,\big\rangle\\
& \hspace{4.7cm}\times\big(\Phi^{-j_3-1}(\mu_3|z_1) 
+({\rm descendants})\big)\,.
\end{aligned}
\end{equation}
The descendant contributions are determined by the 
$(\widehat{\mathfrak s\mathfrak l}_2)_k
\times(\widehat{\mathfrak s\mathfrak l}_2)_k$ current algebra symmetry.
The construction of $n$-point functions is thereby in principle reduced to the
construction of two- and three-point functions:
\begin{align}
\langle\Phi^{j_2}(\mu_2|z_2) 
\Phi^{j_1}(\mu_1|z_1)\rangle
\,=\,& |z_2-z_1|^{-4\Delta_{j_1}} |\mu_1|^2\delta^{(2)}(\mu_2+\mu_1) \nn\\
&\times \big( 
\delta(j_2+j_1+1)+R^{\rmh\sst}(j_1)
\delta(j_2-j_1)\big)
\label{H-2pt} \\
\langle\Phi^{j_3}(\mu_3|z_3)\cdots\Phi^{j_1}(\mu_1|z_1)\rangle
\,=\,& 
|z_3-z_2|^{-2\Delta_{32}^1}|z_3-z_1|^{-2\Delta_{31}^2}
|z_2-z_1|^{-2\Delta_{21}^3} \nn\\ 
&\times \delta^{(2)}(\mu_3+\mu_2+\mu_1) 
D^{\rmh}\big[\begin{smallmatrix} j_3 & j_2 & j_1 \\[.4ex]
\mu_3 & \mu_2 & \mu_1\end{smallmatrix}
\big]\,C^{\rmh\sst}(j_3,j_2,j_1)
\label{H-3pt}
\end{align}
The structure constant $C^{\rmh}(j_3,j_2,j_1)$ 
has the following expression (whose normalization is discussed in \cite{TeH2}),
\begin{align}
C^{\rmh}(j_3,j_2,j_1)=& -\frac{1}{2\pi^3 b}
\left[\frac{\gamma(b^2)b^{2-2b^2}}{\pi}\right]^{-2-\Sigma j_i} 
\frac{\up'(0)}{\up(-b(j_1+j_2+j_3+1))}\label{hb3pt}\\
&\qquad\times
\frac{\up(-b(2j_1+1))\up(-b(2j_2+1))\up(-b(2j_3+1))}
{\up(-bj_{12}^3)\up(-bj_{13}^2)
\up(-bj_{23}^1)}\,,\nonumber
\end{align}
where we used the notation $j^r_{st}\equiv j_s+j_t-j_r$. The quantity
$\delta^{(2)}(\mu_3+\mu_2+\mu_1)  
D^{\rmh}\big[\begin{smallmatrix} j_3 & j_2 & j_1 \\[.4ex]
\mu_3 & \mu_2 & \mu_1\end{smallmatrix}
\big]$ is the Clebsch-Gordan coefficient for $\SLC$ representations of
spins $j_1,j_2,j_3$ in the $\mu$ basis, which is symmetric with
respect to permutations of the indices $1,2,3$ and invariant under
$\mu$-dilatations. 
Explicitly, we have
 \begin{align}
 D^{\rmh} \big[\begin{smallmatrix} j_3 & j_2 & j_1 \\[.4ex]
\mu_3 & \mu_2 & \mu_1\end{smallmatrix}
\big]\,=\,& \pi 
 |\mu_2|^{-2j_1-2j_3-2}|\mu_1|^{2j_1+2}|\mu_3|^{2j_3+2}  
\label{fth3}
 \\ 
  \times & \left[
  \frac{\gamma(j_{23}^1+1)\gamma(j_{13}^2+1) }{\gamma(-j_{123}-1) 
    \gamma(2j_3+2)} \FF\big(j_{123}+2,j_{13}^2+1,2j_3+2,1+
\frac{\mu_1}{\mu_2}\big)
\right.\nn\\
 & \left.  + 
\left|1+\frac{\mu_1}{\mu_2}\right|^{-2(2j_3+1)}
\frac{\gamma(j_{12}^3+1)}{\gamma(-2j_3)}
\FF\big(-j_{23}^1,j_{12}^3+1,-2j_3,1+\frac{\mu_1}{\mu_2}\big) \right] \,,
\nn\end{align}
where we used the notations $j_{rst}=j_r+j_s+j_t$ and 
\begin{equation} \label{notff}
\FF(a,b,c,z)=F(a,b,c,z)F(a,b,c,\bar{z})\,.
\end{equation}
The reflection
coefficient $R^{\rmh\sst}(j)$ is
\begin{align}
R^{\rmh} (j)\,=\,& -
\left(\frac{1}{\pi}b^2\gamma(b^2)\right)^{-(2j+1)}
\frac{\Gamma(+2j+1)}{\Gamma(-2j-1)}
\frac{\Gamma(+b^2(2j+1))}{\Gamma(-b^2(2j+1))}\,.
\label{rhj}
\end{align}
This coefficient is involved in the reflection relation 
\bea
\Phi^j(\mu|z)=R^{\rmh} (j) \Phi^{-j-1}(\mu|z) \,.
\eea
Consistency of the conformal field theory characterized by
these data 
was proven in \cite{TeH3}. There
is good evidence that this conformal field theory
is a quantization of the the classical theory described by the
action (\ref{H-act}).

\subsubsection{The KZ equations}

Due to the $(\widehat{\mathfrak s\mathfrak l}_2)_k
\times(\widehat{\mathfrak s\mathfrak l}_2)_k$ symmetry of the $\H$ model,
the correlation functions
$\Omega^{\rmh\sst}_n$ may alternatively be characterized
as particular solutions to the Knizhnik--Zamolodchikov (KZ) system of 
partial differential equations 
which are subject to certain asymptotic conditions. 
The KZ equations for the case at hand may be written as
\begin{equation}\label{KZ} 
\begin{aligned}
(k-2)\frac{\partial}{\partial z_r}\Omega^{\rmh\sst}_n\,=\,&
{\mathsf H}_r^{}\,\Omega^{\rmh\sst}_n\,,\qquad
{\mathsf H}_r^{}\,\equiv\,&\sum_{s\neq r}\frac{{\rm D}_{rs}}{z_r-z_s},\\
(k-2)\frac{\partial}{\partial \bar{z}_r}\Omega^{\rmh\sst}_n\,=\,&
\overline{\mathsf H}_r^{}\,\Omega^{\rmh\sst}_n,\qquad
\overline{\mathsf H}_r^{}\,\equiv\,&\sum_{s\neq r}\frac{
\overline{\rm D}_{rs}}{\bar{z}_r-\bar{z}_s},\end{aligned}
\end{equation}
where the differential operator ${\rm D}_{rs}$ is defined as
\begin{align}\label{kzmu}
{\rm D}_{rs}\,\equiv\,
{\rm D}^0_r {\rm D}^0_s - & \half \left( {\rm D}^+_r {\rm D}^-_s
+{\rm D}^-_r {\rm D}^+_s \right)
\\
& =\,
\mu_r\mu_s\left[-\left({\partial_{\mu_r}}
-{\partial_{\mu_s}}\right)^2+ \frac{j_r(j_r+1)}{\mu_r^2}+
\frac{j_s(j_s+1)}{\mu_s^2} \right]  \,,
\nonumber
\end{align}
while $\overline{\rm D}_{rs}$ is the complex conjugate of ${\rm D}_{rs}$.
In addition to the equations (\ref{KZ}) we shall consider 
the corresponding complex conjugate equations. 

For later use, let us explain more precisely how the KZ equations
characterize the $\H$ correlators. The main point is that these
equations are first order in $\pp{z_1} $. Thus, the correlator
$\Omega^{\rmh\sst}_n$ is characterized by its behaviour for $z_1\rar
z_2$. Using the OPE, this behaviour is determined by lower correlators
$\Omega^{\rmh\sst}_{n-1}$. These remarks can be expressed more
generally as the following Lemma : 
\begin{lem}\label{baslem} 
Let $\Theta^{\rmh\sst}_{n-1}(M'|Z'|J')$ be a solution of the
system of KZ equations with variables 
\[
M'=(\mu_n,\dots,\mu_3,-\mu_1-\mu_2),\quad Z'=(z_n,\dots,z_3,z_1), \quad
J'=(j_n,\dots,j_3,-j-1),
\]
and let the functions $D(j_3|j_2,j_1)$ 
be arbitrary. 
Then there exists a unique solution $\Theta_n(M|Z|J)$,
\[
M=(\mu_n,\dots,\mu_1),\quad Z=(z_n,\dots,z_1), \quad
J=(j_n,\dots,j_1),
\]
to the KZ-equations (\ref{KZ}) with the
asymptotic behavior
\begin{equation}\label{Corr-asym}
\begin{aligned}
\Theta_n(X|Z|J)=& \int_{-\half+i\R_+}dj_{21}\;
D(j_{21}|j_2,j_1)\,
|z_2-z_1|^{2(\Delta_{j_{21}}-\Delta_{j_2}-\Delta_{j_1})}\\
& \hspace{.25cm}\times
D^{\rmh\sst}\big[\begin{smallmatrix} j_{21} & j_2 & j_1 \\[.4ex]
-\mu_1-\mu_2 & \mu_2 & \mu_1\end{smallmatrix}
\big]
\big(\Theta^{\rmh\sst}_{n-1}(M'|Z'|J')
+{\mathcal O}(z_2-z_1)\big)\,.
\end{aligned}
\end{equation}
\end{lem}


\section{$\H$ correlators from Liouville theory } \label{secpr}

The main result of this paper is a relation between a generic
$\H$ correlator on the sphere 
and Liouville correlators on the sphere involving degenerate fields: 
\begin{equation}\label{main}
\boxed{\quad\begin{aligned}
\big\langle \Phi^{j_n} (\mu_n|z_n)  \cdots
\Phi^{j_1}(\mu_1|z_1)& \big\rangle\,
 =  \, \frac{\pi}{2}(-\pi)^{-n}b\, \times \\
\times \,
 \delta^{(2)}({\textstyle\sum_{i=1}^n\mu_i})
\,  |\Theta_n|^2
\, & \big\langle\, V_{\a_n}(z_n)\cdots V_{\a_1}(z_1)
V_{-\frac{1}{2b}}(y_{n-2}) \cdots V_{-\frac{1}{2b}}(y_1)\,\big\rangle\,.
\end{aligned}\quad}
\end{equation}
The function $\Theta_n$ which appears here is defined as
\begin{align}
\Theta_n^{}(z_1,\dots, z_n|  y_1,\dots, y_{n-2},u)\,=\,
u\prod_{r<s\leq n} z_{rs}^{\frac{1}{2b^2}}
\prod_{k<l\leq n-2} y_{kl}^{\frac{1}{2b^2}}
\prod_{r=1}^n\prod_{k=1}^{n-2}(z_r-y_k)^{-\frac{1}{2b^2}}.
\label{thetan}
\end{align}
The relation (\ref{main})
will hold provided that the respective variables
are related as follows:
\begin{enumerate}
\item The variables $\mu_1,\dots,\mu_n$ are related
to $y_1,\dots,y_{n-2},u$ via 
\begin{equation}\label{keyrel}
\boxed{\quad
\sum_{i=1}^n \frac{\mu_i}{t-z_i} =
u\frac{\prod_{j=1}^{n-2}(t-y_j)}{\prod_{i=1}^n (t-z_i)} 
\, .\quad}
\end{equation}
In particular, since $\sum_{i=1}^n \mu_i=0$, we have $u=\sum_{i=1}^n
\mu_i z_i$.
\item The Liouville parameter $b$
is identified with the $\H$ parameter $b^2=(k-2)^{-1}$.
\item The
Liouville bulk coupling is fixed to the value $\mu_L=\frac{b^2}{\pi^2}$.
\item The
Liouville momenta are given by 
\begin{equation}\label{alpha-j}\a_i=b(j_i+1)+\hib.
\end{equation}
\end{enumerate}

The key element in the correspondence (\ref{main}) 
is clearly the change of variables (\ref{keyrel}).
The relation (\ref{keyrel}) can in principle be
solved to express the variables
$y_r$, $r=1,\dots,n-2$ in terms of the $\mu_s$, $s=1,\dots,n$.
The solution will be unique up to permutations of the variables 
$y_r$. This ambiguity will not jeopardize the validity
of (\ref{main}) since the right hand side is symmetric under
permutations of the $y_r$, $r=1,\dots,n-2$ thanks to the 
mutual locality of the fields $V_{-\frac{1}{2b}}(y_r)$.

\subsection{Proof of the main result}

The proof will be carried out in three steps. The induction on the
number of $\H$ fields $n$ will first be initiated by a check of our main
result in the cases of the two- and three-point $\H$ correlators. The
second step will be to prove the equivalence of the KZ and BPZ equations 
satisfied respectively 
by the $\H$ and Liouville correlators which appear in our main 
result eq. (\ref{main}). This will then be used in the
third step, the inductive proof of our main result for arbitrary $n$. 

\subsubsection{Step 1: Direct proof for the cases $n=2,3$}

The case $n=2$ is completely straightforward, since it reduces to
comparing the
Liouville (\ref{L-2pt}) and $\H$ (\ref{H-2pt})
two-point functions. These turn out to be equivalent due to the
relation between the reflection amplitudes of the models,
\bea
R^{L}(b(j+1)+\hib) = R^{ H} (j) \,.
\eea
Note also the relation between conformal dimensions, which is valid if
$\a=b(j+1)+\hib$ :
\bea
\Delta_j=\Delta_\a+\Delta_{-\hib}+\frac{1}{2b^2}
=\Delta_\a -\frac{k}{4}
\,. \label{HL-dim}
\eea

In the case $n=3$
we start with the Liouville side of the main result (\ref{main}),
which involves a
degenerate Liouville four-point correlator 
\[
\big\langle \,V_{\a_3}(z_3) V_{-\frac{1}{2b}}(y_1)
V_{\a_2}(z_2)V_{\a_1}(z_1)\,\big\rangle \,.
\] 
This can be computed by using 
the operator product expansion involving a degenerate field (\ref{degOPE}).
The result can be written as
\begin{align}\label{degfour}
\big\langle \,  V_{\a_3}(z_3) &  V_{-\frac{1}{2b}}(y_1)
V_{\a_2}(z_2)V_{\a_1}(z_1)\,\big\rangle
\,=\,\\  =\,& 
|z_3-z_2|^{2(\Delta_{\a_1}-\Delta_{\a_2}-\Delta_{\a_3}+\Delta_{-\frac{1}{2b}})}
|z_3-z_1|^{2(\Delta_{\a_2}-\Delta_{\a_1}-\Delta_{\a_3}+\Delta_{-\frac{1}{2b}})}
\nn \\
\times &
|z_2-z_1|^{2(\Delta_{\a_3}-\Delta_{\a_1}-\Delta_{\a_2}-\Delta_{-\frac{1}{2b}})}
|z_3-y_1|^{-4\Delta_{-\frac{1}{2b}}}|1-z|^{2(j_1+1)+b^{-2}}\nn\\
\times & \sum_{\eta=\pm }\, C^{\rml\sst}_{\eta}(\a_3)
C^{\rml\sst}(\a_2,\a_1,\a_3-\frac{\eta}{2b})\,
|z|^{2(\Delta_{\a_3-\frac{\eta}{2b}}-\Delta_{\a_3}+\Delta_{-\frac{1}{2b}})}
\nn \\[-1ex]
&\hspace{1.5cm}\times
\FF(-j_3^\eta+j_1+j_2+1,-j_3^\eta+j_1-j_2,-2j_3^\eta,z).\nn
\end{align}
We have (partially) expressed the variables $\a_i$
in terms of the $j_i$ by using (\ref{alpha-j}), and otherwise used the
notations
\[
z\equiv\frac{(z_1-z_2)(y_1-z_3)}{(z_1-z_3)(y_1-z_2)},\quad
 \quad
j^\eta\equiv \left\{\begin{aligned}
j &  \quad \eta=-, \\ -j-1 &\quad  \eta=+. \end{aligned}
\right. 
\]
We still need to rewrite the Liouville structure constants appearing
in eq. (\ref{degfour}) in terms of $\H$ structure constants. This is
done by using the identities
\begin{align}
C^{\rml\sst}(\a_2,\a_1,\a_3+\frac{1}{2b})C^{\rml\sst}_-(\a_3)=\,&
 -\frac{2\pi^3}{b} 
 \frac{\gamma(j_{12}^3+1)}{\gamma(-2j_3)} C^{\rmh \sst
  }(j_3,j_2,j_1) \,,
\label{3rel}\\
C^{\rml\sst}(\a_2,\a_1,\a_3-\frac{1}{2b})
C^{\rml\sst}_+(\a_3)=\,
& -\frac{2\pi^3}{b} 
 \frac{\gamma(j_{32}^1+1)\gamma(j_{31}^2+1) }{\gamma(-j_{123}-1) 
    \gamma(2j_3+2)} C^{\rmh \sst}(j_3,j_2,j_1) \,.
\label{3rel2}
\end{align}
Furthermore, we can perform the change of variables
(\ref{keyrel}), which leads in particular to the following expression
for $y_1$ and the cross-ratio $z$ :
\begin{equation}
y_1\, =\, -\frac{\mu_1z_2z_3+\mu_2z_3z_1+\mu_3z_1z_2}{\sum_{i=1}^n \mu_iz_i},\quad
z\,=\, 1+\frac{\mu_1}{\mu_2}.
\end{equation}
The two terms $\eta=\pm$ of eq. (\ref{degfour}) thus combine into the
two terms of the $\H$ structure constant $D^H$, equation 
(\ref{fth3})
\begin{multline}
\big\langle \,  V_{\a_3}(z_3)   V_{-\frac{1}{2b}} (y_1)
V_{\a_2}(z_2)V_{\a_1}(z_1)\,\big\rangle\, =\\
=
|z_{32}|^{2(\Delta_{\a_1}-\Delta_{\a_2}-\Delta_{\a_3}-\Delta_{-\frac{1}{2b}})}
|z_{31}|^{2(\Delta_{\a_2}-\Delta_{\a_1}-\Delta_{\a_3}-\Delta_{-\frac{1}{2b}})}
|z_{21}|^{2(\Delta_{\a_3}-\Delta_{\a_1}-\Delta_{\a_2}-\Delta_{-\frac{1}{2b}})}
\\
\times
|u|^{4\Delta_{-\hib}}|\mu_1|^{\frac{1}{b^2}}|\mu_2|^{\frac{1}{b^2}}
|\mu_3|^{ \frac{1}{b^2}}
\times -\frac{2\pi^2}{b}
C^{\rmh \sst}(j_3,j_2,j_1)\,
 D^{\rmh}\big[\begin{smallmatrix} j_3 & j_2 & j_1 \\[.4ex]
\mu_3 & \mu_2 & \mu_1\end{smallmatrix}
\big] \,.
\end{multline}
This has to be multiplied by
\begin{align}
\left|\Theta_3( z_1,z_2 ,z_3|y,u)\right|^2  =\, 
|u|^{-4\Delta_{-\frac{1}{2b}}} 
|\mu_1|^{-\frac{1}{b^2}}|\mu_2|^{-\frac{1}{b^2}}
|\mu_3|^{-\frac{1}{b^2}}
\times  |z_{32}|^{-\frac{1}{b^2}}|z_{31}|^{-\frac{1}{b^2}}|z_{21}|^{-\frac{1}{b^2}}
\nn
\end{align}
Taking into account the relation for conformal dimensions (\ref{HL-dim}),
the result is the desired $\H$ three-point function appearing in the
main result (\ref{main}). 

\subsubsection{Step 2: Correspondence between the differential equations}

We want to prove that validity of the BPZ equations for the 
Liouville correlation function appearing in (\ref{main})
is equivalent to the fact that the right hand side of
(\ref{main}) satisfies the KZ equations.
This observation is essentially due to Stoyanovsky \cite{St}.
For the reader's convenience we shall 
explain the proof of this claim\footnote{There are two 
  differences with Stoyanovsky here. 
First, we impose $\sum_{i=1}^n\mu_i=0$.  
This reduces the number of Liouville degenerate fields to
  $n-2$ instead of $n-1$. Second, our normalization of 
  the operators $\Phi^j(\mu)$ eq. (\ref{phimu}) leads to notable
  simplifications, in particular our function $\Theta_n$ does not
  depend on the spins $j_r$.
}. 
To begin with, let us rewrite 
the system of KZ equations (\ref{KZ}) in an equivalent form by introducing an
arbitrary parameter $t$ and taking linear combinations of the KZ
equations :
\begin{equation}\label{KZt}
\sum_{r=1}^n\frac{k-2}{t-z_r}
\left(\frac{\partial}{\partial z_r}-\frac{\Delta_{j_r}}{t-z_r}\right)
\Omega^{\rmh\sst}_n\;=\;{\mathsf S}(t)\,\Omega^{\rmh\sst}_n,
\end{equation}
where ${\mathsf S}(t)$ is defined as
\begin{equation}
{\mathsf S}(t)=\sum_{r=1}^n\left(\frac{{\mathsf H}_r}{t-z_r}+
\frac{j_r(j_r+1)}{(t-z_r)^2}\right)\,=\,
({\mathsf J}^0(t))^2+\partial_t{\mathsf J}^0(t)
-{\mathsf J}^-(t){\mathsf J}^+(t)\,.
\end{equation}
We have written the result in terms of
 ${\mathsf J}^a(t)=\sum_{r=1}^n \frac{{\rm D}^a_r}{t-z_r} $ because we
 are now interested in the case $t=y_a$ where we have ${\mathsf
 J}^-(y_a)=0$ due to the relation (\ref{keyrel}). This relation also
 implies 
\[
\frac{\partial}{\partial y_a}\,=\,
\sum_{r=1}^n\frac{\mu_r}{y_a-z_r}\frac{\partial}{\partial \mu_r}\,= \, -{\mathsf
 J}^0(y_a) 
\,,
\]
and thus we have 
\bea
 {\mathsf S}(t)\big|_{t=y_a}=\left(\pp{y_a} \right)^2 \,. \label{sepofvar}
\eea
This identity is a nice observation originally due to Sklyanin. 
It implies that the eigenvalue problem for the 
Gaudin Hamiltonians ${\mathsf H}_r\Psi=E_r\Psi$ is equivalent to
the system of $n-2$ separated equations
\begin{equation}
 \p_{y_a}^2\Psi\,=\,\sum_{r=1}^n
\left(\frac{E_r}{y_a-z_r}+\frac{j_r(j_r+1)}{(y_a-z_r)^2}\right)\Psi\,.
\end{equation}
We note that the equation with index $a$ does not contain any
reference to the variables $y_b$, $b\neq a$. The transition
from the original multidimensional eigenvalue problem to a set of
decoupled one-dimensional problems is called the separation of
variables.

It remains to consider the left hand side of the KZ equations 
in the form (\ref{KZt}), specialized to the values $t=y_a$.
We again use the change of variables
(\ref{keyrel}) in the form
\begin{equation}
\mu_r\,=\,u\frac{\prod_{a=1}^{n-2}(z_r-y_a)}{\prod_{s\neq r}(z_r-z_s)}\,.
\label{mur}
\end{equation}
This allows us to derive 
the identity
\bea
\delta_a \mu_{s}\,=\,0,\quad \delta_a\,\equiv\,
\sum_{r=1}^n\frac{1}{y_a-z_r}
\left(\frac{\partial}{\partial z_r}+\frac{\partial}{\partial y_a}\right)-
\sum_{b\neq a}\frac{1}{y_a-y_b}
\left(\frac{\partial}{\partial y_a}-\frac{\partial}{\partial y_b}\right)
. \label{secondid}
\eea
With the help of relations (\ref{sepofvar}) and (\ref{secondid})
it becomes easy to see that the KZ equations are equivalent to the
system of equations
\begin{equation}
(k-2)\left(
\delta_a-\sum_{r=1}^n\frac{\Delta_{j_r}}{(y_a-z_r)^2}\right)\Omega^H_n\,=\,
\p_{y_a}^2\,\Omega^H_n\,.
\end{equation}
It is then straightforward to check that twisting by the function
$\Theta_n$ (\ref{thetan}) yields the BPZ equations, plus the
worldsheet translation invariance equation
\bea
\sum_{r=1}^n\pp{z_r} +\sum_{a=1}^{n-2}\pp{y_a} =0 \,.
\eea

\subsubsection{Step 3: Generalization to arbitrary $n$}

Let us assume that our main result (\ref{main}) has been proven
for all $n'<n$ (with $n\geq 4$).
We will  show that this implies the validity of (\ref{main}) for $n'=n$. 
Since we now know that both sides in (\ref{main}) satisfy the same
first-order differential equations in $z_r$, it is enough to show that
they are equal in the limit $z_{12}\rar 0$ (see the Lemma
\ref{baslem}). In this limit, the OPE (\ref{H-OPE}) reduces the $\H$ correlator
$\Omega^H_n$ to $(n-1)$-point and 3-point correlators:
\begin{align}  \label{mainlhs}
 \big\langle \Phi^{j_n} (\mu_n|z_n)  \cdots
 \Phi^{j_1}(\mu_1|z_1) \big\rangle\, &
= \, 
\\
= \int_{-\half+i\R_+}dj_{21}\int \frac{d^2\mu_{21}}{|\mu_{21}|^2}\;
& \big\langle \Phi^{j_{21}} (\mu_{21}|\infty) \Phi^{j_2}(\mu_2|z_2)
\Phi^{j_1}(\mu_1|z_1)\big\rangle\label{Thetafactor}\nn\\
\times\big( & \big\langle \Phi^{j_n} (\mu_n|z_n)  \cdots
\Phi^{j_3}(\mu_3|z_3)\Phi^{-j_{21}-1}
(\mu_{21}|z_1) \big\rangle
+{\mathcal O}(z_{21})\big).
\nn\end{align}
We need to compare this with the limit $z_{12}\rar 0$ of the right
hand side of equation (\ref{main}), which is
\bea
\frac{\pi}{2}(-\pi)^{-n}b\ 
\delta^{(2)}({\textstyle\sum_{i=1}^n\mu_i}) \, |\Theta_n|^2
\, \big\langle\, V_{\a_n}(z_n)\cdots V_{\a_1}(z_1)
V_{-\frac{1}{2b}}(y_{n-2}) \cdots V_{-\frac{1}{2b}}(y_1)\,\big\rangle.
\label{mainrhs}
\eea
In order to determine the behaviour of this quantity, we have to study
the behaviour of the $y_a$s. 
Relation
(\ref{keyrel}) implies that in the limit $z_2-z_1\rightarrow 0$
one of the variables $y_a$, henceforth taken to be $y_1$, 
will also approach $z_1$,
\begin{equation}
y_1\,=\,
z_1+(z_2-z_1)\frac{\mu_1}{\mu_1+\mu_2}+{\mathcal O}\big((z_2-z_1)^2\big)
\,.
\end{equation}
The remaining $y_a$s are defined through
\begin{equation}
\frac{\mu_1+\mu_2}{t-z_1}+\sum_{r=3}^n
\frac{\mu_r}{t-z_r}=-\frac{\prod_{a=1}^{n-2}(t-y_a)}
     {(t-z_1)\prod_{r=3}^n (t-z_r)} \sum_{r=1}^n z_r\mu_r \,.
\label{asydef}
\end{equation}
We therefore need to study the asymptotic behavior of the
Liouville correlator 
\[
\big\langle \cdots
V_{-\frac{1}{2b}}(y_1)V_{\a_2}(z_2)V_{\a_1}(z_1)\,\big\rangle
\]
in the limit $z_{12},y_1-z_1\rar 0$. A generalization of the operator product expansion
(\ref{L-OPE}) leads to:
\begin{align}
  \big\langle\,   V_{\a_n}(z_n)  \cdots  V_{\a_1}(z_1) & 
V_{-\frac{1}{2b}}(y_{m}) \cdots V_{-\frac{1}{2b}}(y_1)\,\big\rangle\,=\, 
\label{Omegafactor}\\
 = \frac{1}{2\pi}\int_{\frac{Q}{2}+i\R_+}d\a_{21}^{}\;   & \big\langle \,
V_{Q-\a_{21}}(\infty) \,V_{-\frac{1}{2b}}(y_1)
V_{\a_2}(z_2)V_{\a_1}(z_1)\,\big\rangle
\nn\\
\times\big( & \big\langle\,   V_{\a_n}(z_n) \cdots V_{\a_3}(z_3)
V_{\a_{21}}(z_1)
V_{-\frac{1}{2b}}(y_{m}) \cdots V_{-\frac{1}{2b}}(y_2)\,
\big\rangle +{\mathcal O}(z_{21})\big)\,
\,,\nn
\end{align}
where the operator $V_\a(\infty)$ was defined in eq.  (\ref{inftynot}).

We now study the $\Theta_n$ factors. The following rewriting,
\bea
\Theta_n(z_1,\dots,z_n|\mu_1,\dots,\mu_n)\, = \,
(\sum_{i=1}^n\mu_iz_i)^{1+\frac{n}{2b^2}} \prod_{r=1}^n
\mu_r^{-\frac{1}{2b^2}} \prod_{r<s\leq n} z_{rs}^{-\frac{1}{2b^2}}
\prod_{k<l<\leq n-2} y_{kl}^{\frac{1}{2b^2}} \,,
\eea
makes it easy to derive the asymptotic behaviour,
\begin{align}\label{chifactor}
\Theta_n^{}(&  
z_1,z_2,z_3,\dots, z_n|\mu_1,\mu_2,\mu_3,
\dots, \mu_n)
\,=\, |\mu_1+\mu_2|^{-2}\, \times   \\
&\times \,\Theta_{n-1}^{}(z_1,z_3\dots, z_n|
\mu_1+\mu_2,\mu_3,\dots,\mu_n)
\, \Theta_3\big(z_1,z_2,\infty|\mu_1,\mu_2,-\mu_1-\mu_2)
\big(1+{\mathcal O}(z_{21})\big)\,,
\nn
\end{align}
where $\Theta_3(\dots,z_3=\infty|\dots)$ 
is defined as follows:
\[
\Theta_3(\dots,z_3=\infty|\dots)=\lim_{z_3\rar
  \infty}z_3^{-1-\frac{1}{2b^2}}\Theta_3(\dots,z_3|\dots).
\]

The $z_{12}\rar 0$ behaviour of the Liouville side 
eq. (\ref{mainrhs}) is now fully determined in terms of the behaviours
of $\Theta_n$ eq. (\ref{chifactor}) and 
$\Omega^L_{n|n-2}$ eq. (\ref{Omegafactor}). In order to compare it with
the behaviour of the $\H$ side eq. (\ref{mainlhs}), we only need to
apply the
induction hypothesis eq. (\ref{main}) at levels $n'=3$ and
$n'=n-1$, while performing the change of integration variable
$d\a_{21}=b\ dj_{21}$. 
This shows the validity of (\ref{main}) at level $n$.



\subsection{Comparison with the Fateev-Zamolodchikov correspondence }

The first instance of a relation between KZ and BPZ equations was
found by Fateev and Zamolodchikov in \cite{fz86}.
It relates the 
KZ equation satisfied by a generic four-point $\H$ correlator,
\bea
\Omega^H_4=\langle
\Phi^{j_4}(\infty|\infty)\Phi^{j_3}(1|1)\Phi^{j_2}(x|z)\Phi
^{j_1}(0|0)\rangle\,,
\eea
and the BPZ equation satisfied by a Liouville five-point correlator
with one degenerate field \footnote{The Liouville five-point
  correlator has to be multiplied by a factor 
$|\Theta_{\rm FZ}|^2$
  with 
\bea
\Theta_{\rm FZ}=x^{-\alpha _{1}/b}(1-x)^{-\alpha
_{3}/b}(x-z)^{-\alpha _{2}/b}
z^{-2(b^{2}j_{1}j_{2}-\alpha _{1}\alpha
_{2})} (1-z)^{-2(b^{2}j_{3}j_{2}-\alpha _{3}\alpha _{2})}\, .\eea }
\bea
\Omega^L_{4|1}= \langle V_{\alpha_4}(\infty)V_{\alpha_3}(1)V_{-\frac{1}{2b}}(x)
V_{\alpha_2}(z)V_{\alpha_1}(0) \rangle \,.
\eea
The relation holds provided the Liouville momenta are given in terms
of the $\H$ spins by
\bea \left\{ \begin{array}{rl}
 \alpha_2 =& -b(j_1+j_2+j_3+j_4+1) \,,
\\
2\alpha_i-Q=&b(-j_2+j_1+j_3+j_4-2j_i)\ \ \ (i=1,3,4) \,.
\end{array}\right. 
\eea
This relation was originally found in the context of 
the $SU(2)$-WZNW model, which would correspond to 
negative integer values of $k$ and half-integer spins $j_r$ 
in our notations.

The Fateev-Zamolodchikov KZ-BPZ relation has had several applications. For
instance, it
has been extended by one of us
to a relation between the corresponding physical $\H$ and
Liouville correlators in order to prove the crossing symmetry of the
$\H$ model from the crossing symmetry of Liouville theory
\cite{TeH3}.  
However, the relation studied in \cite{TeH3}
involves complicated $j$-dependent
factors. 
It seems impossible to
generalize the 
correspondences based on 
the Fateev-Zamolodchikov KZ-BPZ relation to 
arbitrary $n$-point correlators.

\subsection{$\H$ correlators in the $m$-basis and the $\SLU$ coset model }

There exists a basis of $\H$ primary fields which is particularly
useful in connection with the $\SLR$ coset model:
\bea
\Phi^j_{m\bar{m}}=\int d^2x\ x^{-j-1+m}\bx^{-j-1+\bar{m}} \Phi^j(x)\,.
\eea
where $m$ and $\bar{m}$ are restricted to the set of values
\bea\label{physval}
m=\frac{n+ip}{2}\,, \quad \bar{m}=\frac{-n+ip}{2}\,, \quad n\in \Z\,,
\quad p
\in \R\,.
\eea
The fields $\Phi^j_{m\bar{m}}$ are then related to the
$\Phi^j(\mu)$ as follows
\bea
\label{phijmm}
\Phi^j_{m\bar{m}}\,=\, N^j_{m\bar{m}} \int
\frac{d^2\mu}{|\mu|^2} \mu^{-m}\bar{\mu}^{-\bar{m}}\ \Phi^j(\mu) \,,
\eea
where the normalization factor $N^j_{m\bar{m}}$ is\footnote{We thank
Yu Nakayama and Gaston Giribet for informing us of a mistake in
earlier versions of this formula.} 
\bea
\label{njmm}
N^j_{m\bar{m}}\,=\,
  \frac{\Gamma(-j+m)}{\Gamma(j+1-\bar{m})}
\eea
As a check, we can derive the $\SLU$ coset model's reflection
 amplitude from the reflection coefficient $R^H(j)$ using this
 formula.

We may use (\ref{phijmm}) in order to translate our main result
(\ref{main}) into a representation for correlation functions
of the fields $\Phi^j_{m\bar{m}}$. The integral over the variables
$\mu_r$ can be transformed into an integral over $u,y_1,\dots,y_{n-2}$
by using
\bea
\label{jacobian}
\frac{d^2\mu_1}{|\mu_1|^2}\dots\frac{d^2\mu_n}{|\mu_n|^2}
\delta^{(2)}({\textstyle\sum_{r=1}^n\mu_r})
=\frac{d^2u}{|u|^4}{d^2 y_1}\dots 
{d^2 y_{n-2}} \frac{\prod_{r\neq s}|z_r-z_s|^2
\prod_{a\neq b}|y_a-y_b|^2}{\prod_{r=1}^n\prod_{a=1}^{n-2}|z_r-y_a|^2}.
\eea
This leads to the following expression for 
correlation functions of the  fields  $\Phi^j_{m\bar{m}}$ in
terms of Liouville correlators:
\begin{align}
 \big\langle\, & \Phi^{j_1}_{m_1\bar{m}_1}(z_1)  \cdots
\Phi^{j_n}_{m_n\bar{m}_n}(z_n) \, \big\rangle\, =  \,
\frac{2\pi^3b}{\pi^{2n}(n-2)!}\ \delta\left(i{\textstyle \footnotesize  \sum_{r=1}^n
}(m_r+\bar{m}_r)\right)\delta_{{ \footnotesize \sum_{r=1}^n} (m-\bar{m}) }
\nn \\ & \qquad\times
\prod_{r<s\leq n}(z_r-z_s)^{m_r+m_s+\frac{k}{2}}
(\bar{z}_r-\bar{z}_s)^{\bar{m}_r+\bar{m}_s+\frac{k}{2}}
\prod_{r=1}^n N^{j_r}_{m_r\bar{m_r}}\label{mmain}\\
& 
\qquad\times  
\int_\C d^2 y_1\cdots d^2y_{n-2}\!\!\!\!
\prod_{a<b\leq n-2}|y_a-y_b|^{k}
\prod_{r=1}^n\prod_{a=1}^{n-2}(z_r-y_a)^{-m_r-\frac{k}{2}}
(\bar{z}_r-\bar{y}_a)^{-\bar{m}_r-\frac{k}{2}}
\nn\\
 & \hspace{6cm}\times\big\langle\, V_{\a_n}(z_n)\cdots V_{\a_1}(z_1)
V_{-\frac{1}{2b}}(y_{n-2}) \cdots V_{-\frac{1}{2b}}(y_1)\,\big\rangle\,
\nn\end{align}
The integrals 
in this formula are absolutely convergent if we
assume that $m$ and $\bar{m}$ take the values (\ref{physval}). The
combinatorial factor $\frac{1}{(n-2)!}$ comes from the invariance of 
$\mu_r$ under permutation of the $y_a$s.

The $\H$ correlators in the $\Phi^j_{m\bar{m}}$ basis have a simple
relation to the winding-number 
preserving correlators of fields creating states 
in the continuous spectrum of the
$\SLU$ coset model. 
It suffices to multiply our formula
(\ref{mmain}) with appropriate free boson correlators. For more
details on this procedure, see \cite{rs03} and references therein.

Now the $\SLU$ coset model has been conjectured to be dual to Sine-Liouville theory by Fateev, 
Zamolodchikov and Zamolodchikov. At the perturbative level, this duality is an identity 
between the correlation functions of the two models, once the appropriate identification
of variables
is performed. Therefore, our result (\ref{mmain}) implies that winding-preserving
Sine-Liouville correlators can 
be expressed in terms of Liouville correlators. It was brought to our attention that 
such a relation was already found 
a long time ago by V. Fateev \cite{F96}, using free field calculations. His relation agrees
with our result up to normalizations. \footnote{Moreover, Fateev also found an expression for 
winding-violating correlators in terms of Liouville correlators. }

\section{Critical level limit}

As an application of our previous results,
we shall now consider the critical level limit $k\downarrow 2$. 
The corresponding limit $b\rightarrow \infty$ of the 
Liouville correlation functions is equivalent to the 
semi-classical limit
$b\rightarrow 0$, as follows from the self-duality of Liouville theory
under $b\rightarrow b^{-1}$. By combining 
known results about the 
semi-classical limit of Liouville theory with 
our correspondence (\ref{main})
between Liouville theory and the $H_3^+$ model 
it becomes possible to characterize the 
critical level asymptotics of the correlation
functions in the $H_3^+$ model rather precisely.

In the following two subsections we will 
analyze the critical level limit
directly within the $H_3^+$ model. In this limit, $\H$ correlators can
be expressed as linear combinations of Gaudin eigenvectors,
but we will not be able to determine which
combinations appear. This will become possible by making use of our
$\H$-Liouville correspondence in subsection \ref{comparison}.
Before this, we will review the relevant properties of Liouville
correlators in the semi-classical limit (subsection \ref{lioukk}).

\subsection{Critical level asymptotics of KZ solutions }\label{Gaudin}

Useful information about the structure of correlation functions of the
$H_3^+$ model in the limit $k\downarrow 2$ can be 
obtained from a quasiclassical analysis of the KZ equations.
In the case of solutions to the KZ equations which take values
in tensor products of 
finite-dimensional representations of ${\mathfrak s\mathfrak l}_2$ 
such an analysis was carried out in \cite{RV}. 
Here, however, we will be interested in solutions which take
values in infinite dimensional representations of the 
zero-mode subalgebra ${\mathfrak s\mathfrak l}_2$.
More specifically, for later
convenience we will be interested in
representations from the {\it complementary} series of $SL(2,{\mathbb C})$, 
which corresponds to real values of the spin, $-1<j<0$.
We will furthermore be interested in solutions to both the 
KZ equations and their complex conjugates.
 
Following \cite{RV}, we will look for a solution  
in the sense of formal power series 
in $k-2$ with leading terms of the form
\begin{equation}\label{KZ-ans}
e^{-\frac{1}{k-2}S(z_1\dots z_n)}
\Psi(\mu_1\dots\mu_n|z_1\dots z_n)\big(1+{\cal O}(k-2)\big).
\end{equation}
Inserting (\ref{KZ-ans}) into the KZ equations (\ref{KZ}), one may easily 
verify that these equations are solved to leading order in $k-2$
provided that 
\begin{align}
{\rm (i)}\quad  {\mathsf H}_r\Psi\,=\,E_r\Psi, \quad & \quad 
\bar{\mathsf H}_r\Psi\,=\,\bar{E}_r\Psi, \label{Gaudproblem} \\
{\rm (ii)}\quad  \partial_{z_r}S\,=\,-E_r, \quad & \quad 
\partial_{\bz_r}S\,=\,-\bar{E}_r. \label{actiondiff}
\end{align}
The first of these two equations is the system of common eigenvector equations
for the set of $2n$ commuting Gaudin Hamiltonians ${\mathsf
  H}_r,\bar{\mathsf H}_r $ \cite{Gaud,Skl,Fr}.
As expressed by equation (\ref{actiondiff}), the $2n$ Gaudin eigenvalues
all derive from a potential 
$S=S(z_1\dots z_n)$.

One should note, however, that in our case it does not seem to be completely
straightforward to prove that the relevant solutions neccessarily
have the form proposed in (\ref{KZ-ans}). The proof of the integrability
conditions for (\ref{actiondiff}), $\partial_{z_r}E_s=\partial_{z_s}E_r$,
given in \cite{RV} requires the finite-dimensionality of the 
relevant representations of ${\mathfrak s\mathfrak l}_2$. Instead of 
further investigating  the possibility to give a direct proof of
(\ref{KZ-ans}) we shall justify this ansatz a posteriori with the 
help of the $H_3^+$-Liouville correspondence. For the moment let us
simply adopt (\ref{KZ-ans}) as a working hypothesis and try to see
how far we get.

Like the Gaudin Hamiltonians, the eigenvalues
$E_r=E_r(z_1\dots z_n)$ and $\bar{E}_r$ depend on the worldsheet coordinates. 
The eigenvalues are not all independent, but rather 
restricted by the equations
\begin{equation}\label{MoebiusEr}
\sum_{r=1}^n z_r^k\big(z_rE_r+(k+1)j_r(j_r+1)\big)=0~~{\rm for}~~
k=-1,0,1,
\end{equation}
which express the invariance under worldsheet M\"obius transformations 
$z_r\rightarrow \frac{az_r+b}{cz_r+d}$.

 
As already observed in the previous section, it is possible to perform the
separation of variables in the Gaudin eigenvalue equations
(\ref{Gaudproblem}) by changing variables from $\mu_1,\dots,\mu_n$ to
$y_a$, $a=1,\dots,n-2$, and applying 
Sklyanin's observation (\ref{sepofvar}). The resulting equations
\begin{equation}\label{sepeveqn}
 \frac{\p^2}{\p{y_a}^2}\Psi \,=\,\sum_{r=1}^n
\left(\frac{E_{r}}{y_a-z_r}+\frac{j_r(j_r+1)}{(y_a-z_r)^2}\right)\Psi\,,
\end{equation}
together with their complex conjugate counterparts 
can be solved in a factorized form,
\begin{equation}\label{prodans}
\Psi =
\prod_{a=1}^{n-2}\psi_a(y_a|z_1\dots
z_n)\prod_{b=1}^{n-2}\bar{\psi}_b(\bar{y}_b|z_1\dots z_n)
\,.
\end{equation}
The most general solution to the separated equations (\ref{sepeveqn})
is a linear combination of factorized solutions of the form
(\ref{prodans}). Since the separated equations are second order, there
are $2^{2(n-2)}$ independent solutions. 

\paragraph{Notations. } In this section,
we omit the dependence on antiholomorphic variables
$\bar{z},\bar{\mu}$. However, for the $y_a$ variables which arise from
Sklyanin's separation of variables, no implicit
dependence on $\bar{y}_a$ should be assumed unless explicitly stated.

\subsection{Critical level asymptotics of $\H$ correlators }\label{critcorr}

Let us now consider the  $k\downarrow 2$ asymptotic behaviour of the
$\H$ model $n$-point correlator $\Omega^H_n$. A priori, this behaviour
is given by a linear combination of terms of the form (\ref{KZ-ans})
for different functions $S^\ell(z_1\dots z_n) $ (and thus different sets of Gaudin
eigenvalues $E_r^\ell,\bar{E}_r^\ell$). In turn, each Gaudin
eigenvector $\Psi^\ell$ is a linear combination of factorized
solutions (\ref{prodans}).

Nevertheless, it is natural to expect that locally in $(z_1\dots z_n)$
one function $S^{\rm \sst min}$ will
dominate the asymptotic behaviour of $\Omega^H_n$. This function will
naturally be the one with smallest real part $\Re S^\ell$. As a
working hypothesis, we also assume
that this $S^{\rm \sst min}$ is unique. This means
\begin{equation}\label{critasomega}
\Omega_n^{H}\,\underset{k\downarrow 2}{\asymp}\,e^{-\frac{1}{k-2}
S^{\rm\sst min}(z_1\dots z_n)}
\Psi(\mu_1\dots\mu_n|z_1\dots z_n),
\end{equation}
where $\Psi$ is an eigenvector of the Gaudin Hamiltonians with 
eigenvalues $E_r^{\rm\sst min}=\p_{z_r}S^{\rm \sst min}$.

Sklyanin's observation allows us to write $\Psi$ as a linear combination
of solutions to the Fuchsian differential equations (\ref{sepeveqn})
which have the factorized form 
(\ref{prodans}). 
Now let us use 
the single-valuedness of 
$\Omega^H_n$ w.r.t. the $y_a$ variables
in order to restrict the form of this linear combination.
We will argue that this requirement
not only restrict the coefficients which can be used to write
$\Psi$ as a linear combination of eigenvectors (\ref{prodans}),
they also turn out to impose severe restrictions on the 
eigenvalues $E_1^{\rm\sst min},\dots, E_n^{\rm\sst min}$ themselves.

Let us focus on the dependence of $\Psi$ w.r.t. some $y\in \{y_1\dots
y_{n-2}\}$. The $y$-dependence of a general solution is given by
\begin{equation}\label{chians}
\chi(y,\bar{y}|z_1\dots z_n)\,=\,
\left(\bar{\psi}_1(\bar{y}),\bar{\psi}_2(\bar{y}) \right)
\cdot K\cdot
\biggl(\begin{matrix} \psi_1(y)\\ 
\psi_2(y)\end{matrix}\biggl)\,, 
\end{equation}
where $\psi_1(y),\psi_2(y)$ are
two linearly independent
solutions of the Fuchsian differential equation (\ref{sepeveqn}),
and $K$ is a $2\times 2$ matrix. 
Single-valuedness
of  $\chi(y,\bar{y})$ constrains the $2\times 2$ monodromy
matrices $M_r$ of $\left(\bar{\psi}_1(\bar{y}),\bar{\psi}_2(\bar{y})
\right)$ around $z_r$,
\begin{equation}\label{monod}
M^{\dagger}_r\cdot K \cdot 
M_r^{\phantom{\dagger}}=K~~{\rm for ~all}~~r=1,\dots,n.
\end{equation}
This is a 
highly overdetermined
system of equations for the matrix $K$, 
which will also restrict the monodromy matrices $M_r$. If, for example,
all $M_r$ happen to be contained in the subgroup $SU(1,1)$ of 
$SL(2,{\mathbb C})$, we may use $K={\rm diag}(1,-1)$ to solve 
(\ref{monod}).
However, for generic systems of Gaudin eigenvalues $E_1,\dots, E_n$, the
system of equations (\ref{monod})
does not have any solution. 
The problem to find eigenvalues
$E_1,\dots, E_n$ which allow one to construct single-valued solutions
of the form (\ref{chians}) does not seem to have a 
direct solution so far. We shall explain in the following that a
solution is provided by the 
uniformization theory of Riemann surfaces with conical singularities.

\subsection{Semiclassical behaviour of Liouville correlators}

\label{lioukk}

We shall now consider the limit $b\rightarrow\infty$ 
of the Liouville correlators $\Omega^L_{n|n-2}$ which are involved in
the correspondence with the $\H$ model, equation (\ref{main}).
 The self-duality of Liouville
theory implies that this limit is equivalent to the semiclassical 
limit $b\rightarrow 0$.
The classical Liouville field
$\vf$ is now recovered from the quantum Liouville field 
$\phi=\frac{1}{2}\partial_{\a}V_\a\big|_{\a=0}$ via $\vf=\frac{2}{b}\phi$
(instead of $\vf=2b\phi$ in the case $b\rar 0$). 
The discussion of the semiclassical limit
of Liouville correlators in \cite{ZZ} may 
be applied to construct the leading asymptotics of Liouville correlation
functions in the standard WKB form $e^{-b^2S^{\rml\sst}[\varphi_{\rm cl}]}$, 
where $S^{\rml\sst}[\varphi_{\rm cl}]$ is the Liouville action
evaluated on a suitable solution of the classical Liouville
equation of motion, 
\begin{equation}\label{LiouEOM}
\partial_y\partial_{\bar{y}}\varphi_{\rm cl}(y,\bar{y})\,=\,
\frac{1}{2}e^{\varphi_{\rm cl}(y,\bar{y})}.
\end{equation}
To be more precise, let us note
that our Liouville correlator $\Omega^{L}_{n|n-2}$ contains insertions
of $n+(n-2)$ exponential fields $V_{\a_k}$ among which the 
$n-2$ fields $V_{-\frac{1}{2b}}$ are ``light'' ($\a\propto \frac{1}{b}$),
whereas the fields $V_{\a_r}$, $r=1,\dots,n$ are ``heavy'' in the
terminology of \cite{ZZ}, since $\a_r=b(j_r+1)+\hib$ with $j_r$
fixed. 
The importance of the distinction between light and heavy fields
becomes clear when considering the OPE between the Liouville
field $\vf$ and a generic primary field $V_{\a}$,
\begin{equation}\label{phiope}
\vf(y,\bar{y})V_{\a}(z,\bar{z})\,
\sim\,-\frac{2\a}{b}\log|y-z|^2+{\mathcal O}(1).
\end{equation}
In the limit $b\rightarrow\infty$, the insertion of a light
field does not influence the solution of the Liouville equation,
whereas insertion of a heavy field implies that the 
classical solution must diverge near the insertion point $z$ 
(this divergence makes it
necessary to regularize the action $S^{\rml\sst}$, see below).
Path integral arguments \cite{ZZ} lead to the following behaviour for
$\Omega^L_{n|n-2}$:
\begin{equation}\label{Lasymp}\begin{aligned}
\Omega^{\rml\sst}_{n|n-2}
\,\underset{k\downarrow 2}{\asymp}
\,\exp\big(\!\!-b^2S^{\rml\sst}[\vf_{\rm cl}]\big)\,
\prod_{a=1}^{n-2} e^{-\frac{1}{2}\varphi_{\rm cl}(y_a,\bar{y}_a)},
\end{aligned}
\end{equation} 
where $S^{\rml\sst}$ and $\vf_{\rm cl}$ are defined more precisely as
follows:
\begin{itemize}
\item
The classical solution $\varphi_{\rm cl}(y,\bar{y}|z_1\dots z_n)$ of the Liouville
equation
(\ref{LiouEOM})
is defined by the boundary conditions
\begin{equation}\label{asympconds}
\begin{aligned}
\varphi_{\rm cl}(y,\bar{y})\,\sim& 
\,-2(j_r+1)\log|y-z_r|^2+{\cal O}(1)~~{\rm for}~~
y\rightarrow z_r,\\
\varphi_{\rm cl}(y,\bar{y})\,\sim& \,-2\log|y|^2+{\cal O}(1)~~
{\rm for}~~ y\rightarrow\infty\,.
\end{aligned}\end{equation}
Existence and uniqueness of a solution to the Liouville equation
is guaranteed by the uniformization theorem \cite{Pi,Tr} for 
the Riemann surface 
\[
\Sigma_n\,\equiv\,
{\mathbb P}^1\setminus\{z_1,\dots,z_n\}\,,
\]
with conical singularities of order $j_r+1$ at the points
$z_1,\dots,z_n$. 
\item
The Liouville action $S^{\rml\sst}[\vf_{\rm cl}]$ is regularized as follows: 
$S^{\rml\sst}[\vf]\equiv
\lim_{\epsilon\rightarrow 0}S^{\rml\sst}_{\epsilon}[\vf]$, where
\begin{equation}
\begin{aligned}
S^{\rml\sst}_{\epsilon}[\vf]\;=\; & \frac{1}{4\pi}
\int_{X^{\epsilon}}d^2z \,
\Big(\,|\partial_z\vf|^2+ e^{\vf}\,\Big)+\bigg(\frac{\epsilon}{2\pi}\int_{\partial D_{n+1}} dx \;\vf-2\log\epsilon\bigg)\\
  - & \sum_{r=1}^{n}
\bigg(\frac{j_r+1}{2\pi\epsilon}\int_{\partial D_r}dx \;
\vf+2(j_r+1)^2\log\epsilon\bigg),
\end{aligned}
\end{equation}
where $D_r=\{z\in{\mathbb C};|z-z_r|<\epsilon\}$, 
$D_{n+1}=\{z\in{\mathbb C};|z|>1/\epsilon\}$, and 
$X^{\epsilon}=D_{n+1}\setminus\bigcup_{r=1}^{n} D_r$.
\end{itemize}

\subsection{Gaudin eigenvalues from accessory parameters}\label{comparison}

Let us now deduce the asymptotic behaviour of the $\H$ correlator
$\Omega^H_n$ from the Liouville theory result (\ref{Lasymp}) through the
$\H$-Liouville correspondence.

First, the behaviour is indeed of the form
(\ref{critasomega}), with only a single potential $S(z_1\dots z_n)$. The 
potential which appears is
\bea
S \, = \, S^L[\varphi_{\rm cl}]\,.
\eea
Moreover, the corresponding eigenvector is 
\bea
\label{compare}
\Psi  \,=\, \delta^{(2)}({\textstyle \sum_{r=1}^n}\mu_r)\ |u|^2\
   \prod_{a=1}^{n-2} e^{-\frac{1}{2}\varphi_{\rm cl}(y_a,\bar{y}_a)}\,.
\eea
Let us now find out which objects from Liouville theory correspond to
   the eigenvalues $E_1\dots E_n$.
First we 
rewrite the Liouville equation in a Fuchsian form, similar to the
Gaudin eigenvalue equations with separated variables
(\ref{sepeveqn}). For this we use the 
classical energy-momentum tensor 
$T_\varphi$ defined as
\begin{equation}\label{Lioufuchs}
(\partial_y^2-T_{\varphi})e^{-\frac{1}{2}\varphi}=0\quad\Leftrightarrow\quad 
T_{\varphi}=-{\textstyle\frac{1}{2}}(\partial_y\varphi)^2+\partial_y^2\varphi
\ .
\end{equation}
The quantity $T_\varphi$ thus defined 
is holomorphic iff $\vf$ satisfies the Liouville equation (\ref{LiouEOM}).
If $\vf$ is furthermore taken to be the unique classical solution 
with singular behavior specified in (\ref{asympconds}) then
its behaviour near the singularities (including $y=\infty$) implies
that 
$T_\vf$ 
can be expressed as
\begin{equation}\label{Texp}
T_{\varphi}(y|z_1\dots z_n)=\sum_{r=1}^n
\left(\frac{-j_r(j_r+1)}{(y-z_r)^2}+\frac{C_r(z_1\dots z_n)}{y-z_r}\right)\,
.
\end{equation}
The functions $C_r$ have become famous under the name of
accessory parameters. 
The regularity of $\vf$ near $y=\infty$ requires the three conditions
\begin{equation}\label{MoebiusEr2}
\sum_{r=1}^n z_r^k\big(z_rC_r-(k+1)j_r(j_r+1)\big)=0~~{\rm for}~~
k=-1,0,1.
\end{equation}
Otherwise
it is difficult to determine the accessory parameters more 
explicitly. 
Nevertheless, they can be shown \cite{CMS,TZ} to be related to the classical
Liouville action of the solution $\varphi_{\rm cl}$ via
\bea
C_r=-\p_{z_r} S^L,\quad \bar{C}_r=-\p_{\bz_r}S^L.
\eea

These observations immediately imply that the function 
$\Psi$ from equation (\ref{compare}) is an eigenvector
for the Gaudin Hamiltonians in separated variables, with 
eigenvalues  
\begin{equation}\label{EvsC}
E_r\,=\,-C_r, \quad \bar{E}_r\,=\, -\bar{C}_r,\quad r=1\dots n \,.
\end{equation}
It seems quite remarkable that the uniqueness of the solution 
$\varphi_{\rm cl}$ to the uniformization problem 
implies the uniqueness of the choice (\ref{EvsC})
for the eigenvalues $E_1,\dots,E_r$, thus uniquely solving
the
``Gaudin single-valued eigenvector problem'' 
that was formulated in subsection \ref{critcorr}. 
This relation is worth further
investigation. In particular, we would like to understand
its representation-theoretic origins 
in more detail. This seems to require some generalization
of the discussion in \cite{Fr}.

\acknowledgments{We are grateful to Volker Schomerus for interesting
conversations. S. R. is grateful to Vladimir Fateev for pointing and discussing 
the unpublished note \cite{F96}. 
We acknowledge support by the EUCLID European network,
  contract number HPRN-CT-2002-00325, and also in part by the PPARC
  rolling grant PPA/G/O/2002/00475. S.R. wishes to thank the Freie
  Universit\"at Berlin for hospitality. J.T. is grateful for
  support from the Deutsche Forschungsgemeinschaft (DFG) via a 
  Heisenberg fellowship. 
Both authors are grateful to SPhT, Saclay for
  hospitality. 
}


\renewcommand{\thesection}{A}
\appendix

\appendix

\section{The $\H$ model with the standard $x$ variables \label{H-xvar}}

In our review of the $\H$ model in the main text, the standard basis
$\Phi^j(x)$ of fields has been replaced with the new basis,
\bea
\Phi^j(\mu|z)=\frac{1}{\pi}
|\mu|^{2j+2} \int_\C d^2x\ e^{\mu x-\bar{\mu}\bar{x}} \Phi^j(x|z) \,,
\label{phimu2}
\eea
which is more convenient for our correspondence with
Liouville theory. 
In this Appendix we explain how to derive the structure
constants of the model with $\mu$ variables from the well-known expressions
with $x$ variables. 

The OPE of the $\H$ model in the $x$ variables reads
\begin{equation}\label{H-OPEx}
\begin{aligned}
\Phi^{j_2}(x_2|z_2) & 
\Phi^{j_1}(x_1|z_1)\,=\,\\ 
& =\,\int_{-\half+i\R_+}dj_3\int d^2 x_3 \;
\big\langle\, \Phi^{j_3}(x_3|\infty) \Phi^{j_2}(x_2|z_2-z_1)
\Phi^{j_1}(x_1|0) \,\big\rangle\\
& \hspace{4.1cm}\times\big(\Phi^{-j_3-1}(x_3|z_1) 
+({\rm descendants})\big)\,.
\end{aligned}
\end{equation}
The construction of $n$-point functions is thereby reduced to the
construction of two- and three-point functions:
\begin{align}
\langle\Phi^{j_2}(x_2|z_2) 
\Phi^{j_1}(x_1|z_1)\rangle
\,=\,& |z_2-z_1|^{-4\Delta_j} \big( 
\delta^{(2)}(x_2-x_1)\delta(j_2+j_1+1)+\nn\\
&\hspace{1.3cm}+B^{\rmh\sst}(j_1)|x_2-x_1|^{4j_1}
\delta(j_2-j_1)\big)\\
\langle\Phi^{j_3}(x_3|z_3)\cdots\Phi^{j_1}(x_1|z_1)\rangle
\,=\,& 
|z_3-z_2|^{2\Delta_{32}^1}|z_3-z_1|^{2\Delta_{31}^2}
|z_2-z_1|^{2\Delta_{21}^3} \nn\\ 
&|x_3-x_2|^{2j_{32}^1}|x_3-x_1|^{2j_{31}^2}
|x_2-x_1|^{2j_{21}^3}\nn\\
&\times 
C^{\rmh}(j_3,j_2,j_1)\,, \label{H-3ptx}
\end{align}
where the structure constant $C^{\rmh}(j_3,j_2,j_1)$ has been given
in equation (\ref{hb3pt}), and the constant $B^{\rmh \sst}(j)$ is 
\bea
 B^{\rmh}(j)=\frac{1}{\pi b^2} \frac{(-\pi
b^2\gamma(-b^2))^{2j+1}}{\gamma(-b^2(2j+1))}\,. 
\eea
The reflection coefficient $R^H(j)$ (\ref{rhj}) 
can be related to the
constant $B^{\rmh \sst}(j)$ by applying the Fourier transform to the
two-point function of the model. The relation is:
\bea
\label{relRB}
R^H(j)=B^{\rmh \sst}(j) \times |\mu|^{4j+2}\int_\C d^2x\
e^{\mu x-\bar{\mu}\bar{x}}|x|^{4j}=B^{\rmh \sst}(j) \times
\pi\gamma(2j+1) \,.
\eea
Note that the integral over $x$ diverges. Thus this integral 
has to be regularized or interpreted
in terms of distributions.

The structure constant $D^H$ (\ref{fth3}) can similarly be related to the
$x$-dependent factors in the three-point function eq. (\ref{H-3ptx}):
\begin{align}\label{DHdef}
\delta^{(2)}(&\mu_1+\mu_2+\mu_3) 
 D^{\rmh}\big[\begin{smallmatrix} j_3 & j_2 & j_1 \\[.4ex]
\mu_3 & \mu_2 & \mu_1\end{smallmatrix}
\big]
\\
=\,&  \frac{1}{\pi^3} \prod_{i=1}^3|\mu_i|^{2j_i+2}
\int_\C d^2x_3d^2x_2 d^2x_1\ e^{\sum_r 
(\mu_rx_r-\bar{\mu}_r\bar{x}_r)}
|x_{12}|^{2j_{12}^3}|x_{13}|^{2j_{13}^2}|x_{23}|^{2j_{23}^1} \,.
\nn
\end{align}
The computation can be performed as follows: first change $x_{1,2}\rar
x_{1,2}+x_3$ so that the integral over $x_3$ can be performed and
yields a $\pi^2\ \delta^{(2)}(\mu_1+\mu_2+\mu_3)$ prefactor. Then, perform
the changes of variables $x_2=\lambda x_1$ and then
$x_1\rar x_1/(\mu_1+\lambda\mu_2)$ to obtain
\begin{multline}
D^H= \frac{1}{\pi}\prod_{i=1}^3|\mu_i|^{2j_i+2} \int d^2x_1\
e^{\mu_1 x_1-\bar{\mu}_1 \bar{x}_1}
|x_1|^{2(\sum j +1)} 
\\
\times \int d^2\lambda\
|\mu_1+\lambda\mu_2|^{-2(\sum j+2)} |\lambda|^{2j^1_{23}}
|1-\lambda|^{2j_{12}^3} \,.
\end{multline}
The first integral has already been used in equation (\ref{relRB})
above. The integral over $\lambda$ was computed by Dotsenko
\cite{dotfr} in the following form:
\begin{multline}
\frac{1}{\pi}\int_\C d^2x\ |x|^{2a}|1-x|^{2b}|x-y|^{2c} 
\\
=\frac{\gamma(b+1)\gamma(-c-a-b-1)}{ \gamma(-a-c)}
\FF(-c,-c-1-a-b,-c-a,y)
\\
+\frac{\gamma(a+1)
\gamma(c+1)}{\gamma(a+c+2)
}|y|^{2(a+c+1)}\FF(-b,a+1,a+c+2,y) \,,
\label{bigint}
\end{multline}
where the notation $\FF$ was defined in equation (\ref{notff}).

\section{Special functions \label{apup} }

The function $\gamma(x)$ is built from Euler's Gamma function:
\bea
\gamma(x)=\frac{\Gamma(x)}{\Gamma(1-x)}\,.
\eea
The function $\up$ is defined for $0<\Re x <Q$ by
\bea
\text{log}\up=\int_{0}^{\infty}\frac{dt}{t}\left\lbrack
\left(\frac{Q}{2}-x\right)^{2}e^{-t}-
\frac{\text{sinh}^{2}(\frac{Q}{2}-x)\frac{t}{2}}{\text{sinh}\frac{bt}{2}
\text{sinh}\frac{t}{2b}}\right\rbrack \,.
\eea
This function can be extended to a holomorphic function on the complex
plane thanks to the shift equations
\bea
\up(x+b)=\gamma(bx)b^{1-2bx}\up(x)\, ,\quad
\up(x+1/b)=\gamma(x/b)b^{2x/b-1}\up(x)\,.
\eea




\begin{thebibliography}{99}


\bibitem{Pol} A.M. Polyakov: {\it Quantum gravity in two dimensions},
  Mod. Phys. Lett. {\bf A2} (1987) 893-898

\bibitem{KPZ} V.G. Knizhnik, A.M. Polyakov, A.B. Zamolodchikov,
   {\it Fractal structure of 2d-quantum gravity}, 
   Mod. Phys. Lett. {\bf A3} (1988) 819-826

\bibitem{FWBFR} P. Forgacs, A. Wipf, J. Balog, L. Feher, 
   L. O'Raifeartaigh, {\it Liouville and Toda theories as conformally reduced 
   WZNW theories}, Phys. Lett. {\bf B227} (1989) 214

\bibitem{TR} J. Teschner: {\it Liouville theory revisited.}
  {\it Class.\ Quant.\ Grav.} {\bf 18} (2001)  R153-R222

\bibitem{TT} J. Teschner: {\it A lecture on the Liouville 
  vertex operators}, Int. J. of Mod. Phys. {\bf A \bf 19},\\ 
  Supplement (2004) 436-458

\bibitem{Na}  Yu. Nakayama: {\it Liouville Field Theory -- 
  A decade after the revolution}, Int. J. Mod. Phys. 
  {\bf A19} (2004) 2771-2930

\bibitem{Ga} K. Gawedzki: {\it Non-compact WZW conformal field
theories}, in: J. Fr\"ohlich et al. (Eds.), Proceedings of NATO ASI
Cargese 1991, Plenum, 1992, pp. 247-274

\bibitem{TeH1} J. Teschner: {\it On structure constants and fusion rules
   in the $SL(2,\C)/SU(2)$-WZNW model,}
   Nucl. Phys. {\bf B546} {(1999)} {390-422}

\bibitem{TeH2} J. Teschner: 
   {\it Operator product expansion and factorization in the 
   $H_3^+$-WZNW model,} Nucl. Phys. {\bf B571} {(2000)} {555-582}

\bibitem{TeH3} J. Teschner: 
   {\it Crossing symmetry in the $H_3^+$ WZNW model,} 
   Phys.\ Lett.\ B {\bf 521} (2001) 127

\bibitem{MO}  J. Maldacena, H. Ooguri:
  {\it Strings in $AdS_3$ and the $SL(2,R)$ WZW Model,} 
  I:   J.Math.Phys. {\bf 42} (2001) 2929-2960, 
  II:  J.Math.Phys. {\bf 42} (2001) 2961-2977,
  III: Phys.Rev. {\bf D64} (2001) 106005.

\bibitem{HK} K. Hori, A. Kapustin: {\it Duality of the Fermionic 
    2d Black Hole and N=2 Liouville Theory as Mirror Symmetry},
    JHEP {\bf 0108} (2001) 045

\bibitem{GK}  A. Giveon, D. Kutasov: {\it Little String Theory in a
    Double Scaling Limit}, JHEP 9910 (1999) 034;\\ 
 {\it Comments on Double Scaled Little String Theory}, JHEP 0001 (2000) 023




\bibitem{CF} P. Christe, R. Flume:
{\it  On the identification of finite operator algebras in
two-dimensional conformally invariant field theories},
Phys.Lett. {\bf B188} (1987) 219

\bibitem{FGPP}
P. Furlan, A. Ch. Ganchev, R. Paunov, V. B. Petkova:
{\it Solutions of the Knizhnik - Zamolodchikov Equation
with Rational Isospins and the Reduction to the Minimal Models},
Nucl.Phys. {\bf B394} (1993) 665-706;
A.Ch. Ganchev, V.B. Petkova:
{\it Reduction of the Knizhnik-Zamolodchikov Equation
- a Way of Producing Virasoro Algebra Singular Vectors,}
Phys. Lett. {\bf B293} (1992) 56-66;
 P. Furlan, A.Ch. Ganchev, V.B. Petkova:
{\it $A_1^{(1)}$ Admissible Representations -
Fusion Transformations and Local Correlators},
Nucl.Phys. {\bf B491} (1997) 635-658

\bibitem{PRY} J.L. Petersen, J. Rasmussen, M. Yu:
{\it Hamiltonian Reduction of $SL(2)$-theories at the Level of Correlators},
Nucl. Phys. B457 (1995) 343

\bibitem{B} H.M. Babujian:
{\it Off-Shell Bethe Ansatz Equation and N-point Correlators in
the SU(2) WZNW Theory},
J.Phys. A26 (1993) 6981-6990

\bibitem{BF} H.M. Babujian, R. Flume:
{\it Off-Shell Bethe Ansatz Equation for Gaudin Magnets and
Solutions of Knizhnik-Zamolodchikov Equations},
Mod.Phys.Lett. {\bf A9} (1994) 2029-2040

\bibitem{FFR} B. Feigin, E. Frenkel, N. Reshetikhin:
{\it Gaudin Model, Bethe Ansatz and Critical Level},
Commun.Math.Phys. {\bf 166} (1994) 27-62



\bibitem{Fr} E. Frenkel, {\it Affine algebras, Langlands duality and 
  Bethe ansatz.} in: Iagolnitzer, Daniel (ed.), Proceedings of the XIth 
  ICMP, Paris, 1994. Cambridge, MA: International Press. 606-642 (1995)
  [arXiv:q-alg/9506003]


\bibitem{DO}  H. Dorn, H.-J. Otto: {\it Two and three-point functions in 
   Liouville...}, Nucl.Phys. {\bf B429} (1994) 375

\bibitem{ZZ} A. Zamolodchikov, Al. Zamolodchikov: {\it 
  Structure Constants and Conformal Bootstrap in Liouville Field Theory},
  Nucl.Phys. {\bf B477} (1996) 577-605

\bibitem{Th} C.B. Thorn: {\it Liouville Perturbation Theory},
   Phys.Rev. {\bf D66} (2002) 027702

\bibitem{St}
A.~V. Stoyanovsky, {\it A relation between the knizhnik--zamolodchikov and
  belavin- -polyakov--zamolodchikov systems of partial differential equations},
   \href{http://arXiv.org/abs/math-ph/0012013}{{\tt math-ph/0012013} v3}.

\bibitem{fz86}
A.~B. Zamolodchikov and V.~A. Fateev, {\it Operator algebra and correlation
  functions in the two- dimensional wess-zumino su(2) x su(2) chiral model},
  {\em Sov. J. Nucl. Phys.} {\bf 43} (1986) 657--664.


\bibitem{rs03}
S.~Ribault and V.~Schomerus, {\it Branes in the 2-d black hole},  {\em JHEP}
  {\bf 02} (2004) 019 [\href{http://arXiv.org/abs/hep-th/0310024}{{\tt
  hep-th/0310024}}].

\bibitem{F96} V. Fateev, {\it Relation between Sine-Liouville and Liouville correlation 
functions}, unpublished note.

\bibitem{RV} N.Reshetikhin, A.Varchenko, {\it Quasiclassical 
     asymptotics of solutions to the KZ equations,} 
     Geometry, Topology and Physics for R. Bott, 
     Intern. Press (1995), 293-322 [arXiv:hep-th/9402126]

\bibitem{Gaud} M. Gaudin, {\it Diagonalization d'une class d'hamiltoniens de spin}
J. Physique {\bf 37} (1976) 1087-1098

\bibitem{Skl} E.K.~Sklyanin, {\it Separation of variables in the Gaudin model},
J. Sov. Math. {\bf 47} (1989) 2473-2488

\bibitem{Pi} E.~Picard, {\it De l'equation $\Delta u=k e^u$ sur une surface
de Riemann ferm\'{e}e}, J. Math. Pure Appl. (4) \textbf{9} (1893),
273-291, and: {\it De l'int\'{e}gration de l'\'{e}quation $\Delta u = e^u$
sur une surface de Riemann ferm\'{e}e}, Crelle's J. \textbf{130}
(1905) 243

\bibitem{Tr} M.~Troyanov, {\it Prescribing curvature on compact surfaces with conical
singularities}, Trans. of the Amer. Math. Soc. \textbf{134} (1991),
793-821




\bibitem{CMS}  L. Cantini, P. Menotti, D. Seminara, 
  {\it Liouville theory, accessory parameters and 2+1 dimensional gravity},  
   Nucl. Phys. {\bf B638} (2002) 351-377

\bibitem{TZ}  L. Takhtajan, P. Zograf,
  {\it Hyperbolic 2-spheres with conical singularities, 
   accessory parameters and Kaehler metrics on $\mathcal{M}_{0,n}$},
   Trans. Amer. Math. Soc. {\bf 355} (2003) 1857--1867


\bibitem{dotfr}
V.~S. Dotsenko, {\it S\'erie de cours sur la th\'eorie conforme},
  http://www.lpthe.jussieu.fr/DEA/dotsenko.html,\ $\sim$\! 1998. 

\end{thebibliography}


\providecommand{\href}[2]{#2}\begingroup\raggedright

\endgroup

\end{document}